\documentclass[a4paper,11pt]{article}

\pdfoutput = 1 

\usepackage{jheppub} 

\usepackage[utf8]{inputenc}
\usepackage{lmodern}
\usepackage[T1]{fontenc}
\usepackage{textcomp}

\usepackage{amsfonts}
\usepackage{amsmath}
\usepackage{amssymb}
\usepackage{units}
\usepackage{slashed}
\usepackage{array}

\usepackage{caption}
\usepackage
{subcaption}
\usepackage{tabularx}
\usepackage{booktabs}
\usepackage{multirow}
\usepackage{xcolor}

\usepackage{tikzexternal}

\usepackage{dcolumn}
\newcolumntype{d}[1]{D{.}{.}{#1}}

\newcommand{\Width}{ \smaller * \textwidth } 
\newcommand{\Height}{ \smaller * \Width } 

\usepackage{environ}
\makeatletter
\environfinalcode{}
\NewEnviron{xsubcaption}[1][-1sp]
 {\let\xsc@theoptcaption\@empty
  \let\xsc@thecaption\@empty
  \let\xsc@thelabel\@empty
  \expandafter\xsc@checkcaption\BODY\caption\xsc@checkcaption
  \expandafter\xsc@checklabel\BODY\label\xsc@checklabel
  \begingroup\edef\x{\endgroup
    \noexpand\subcaptionbox
      \ifx\xsc@theoptcaption\@empty\else
        [\unexpanded\expandafter{\xsc@theoptcaption}]%
      \fi
      {\unexpanded\expandafter{\xsc@thecaption}%
       \ifx\xsc@thelabel\@empty\else
         \noexpand\label{\unexpanded\expandafter{\xsc@thelabel}}%
       \fi}
      \ifdim#1=-1sp 
      \else
        [#1]%
      \fi
      {\unexpanded{\renewcommand\caption[2][]{}\renewcommand{\label}[1]{}}%
       \unexpanded\expandafter{\BODY}}}\x
}

\long\def\xsc@checkcaption#1\caption#2\xsc@checkcaption{%
  \if\relax\detokenize{#2}\relax
    \expandafter\@gobble
  \else
    \expandafter\@firstofone
  \fi
  {\expandafter\xsc@getcaption\BODY\xsc@getcaption}%
}
\long\def\xsc@getcaption#1\caption#2#3\xsc@getcaption{%
  \if[\detokenize{#2}%
    \expandafter\@firstoftwo
  \else
    \expandafter\@secondoftwo
  \fi
  {\expandafter\xsc@getcaptionopt\BODY\xsc@getcaptionopt}%
  {\def\xsc@thecaption{#2}}%
}
\long\def\xsc@getcaptionopt#1\caption[#2]#3#4\xsc@getcaptionopt{%
  \def\xsc@theoptcaption{#2}%
  \def\xsc@thecaption{#3}%
}
\long\def\xsc@checklabel#1\label#2\xsc@checklabel{%
  \if\relax\detokenize{#2}\relax
    \expandafter\@gobble
  \else
    \expandafter\@firstofone
  \fi
  {\expandafter\xsc@getlabel\BODY\xsc@getlabel}%
}
\long\def\xsc@getlabel#1\label#2#3\xsc@getlabel{%
  \def\xsc@thelabel{#2}%
}
\makeatother

\makeatletter
\renewcommand*\env@matrix[1][c]{\hskip -\arraycolsep
  \let\@ifnextchar\new@ifnextchar
  \array{*\c@MaxMatrixCols #1}}
\makeatother

\newcommand{\software}[1]{\texttt{\uppercase{#1}}}
\newcommand{\onehalf}{{\nicefrac{1}{2}}}
\newcommand{\threehalf}{{\nicefrac{3}{2}}}
\newcommand{\diag}{\operatorname{diag}}
\newcommand{\abs}[1]{\left|#1\right|}

\hypersetup{
  unicode = true 
, pdftitle = {Long-lived higgsinos as probes of gravitino dark matter at the LHC} 
, pdfauthor = {S. Bobrovskyi, J. Hajer,S. Rydbeck} 
, breaklinks = true 
, pdfdisplaydoctitle = true 
}

\allowdisplaybreaks[1] 

\title{
\vspace{-4.5ex}
{\normalsize \rm \raggedright
DESY 12--175\\[10ex]
}
Long-lived higgsinos as probes of gravitino dark matter at the LHC
}

\author{S. Bobrovskyi,}
\author{J. Hajer}
\author{and S. Rydbeck}

\affiliation{Deutsches Elektronen-Synchrotron DESY, Notkestra\ss e 85, D-22607 Hamburg, Germany}

\emailAdd{sergei.bobrovskyi@desy.de}
\emailAdd{jan.hajer@desy.de}
\emailAdd{sara.rydbeck@desy.de}

\abstract{%
We investigate the LHC sensitivity to supersymmetric models with light higgsinos, small R-parity breaking and gravitino dark matter.
The limits on decaying gravitino dark matter from gamma-ray searches with the Fermi-LAT put a lower bound on the higgsino-like neutralino NLSP decay length, giving rise to a displaced-vertex collider signature.
Using publicly available tools for simulation of signal, background and detector response, we find that higgsinos with masses of \unit[100 -- 400]{GeV} and R-parity violation of $\zeta \sim 10^{-8} \text{ -- } 10^{-9}$ can show up in the \unit[8]{TeV} LHC data with $\unit[10 \text{ -- } 30]{fb^{-1}}$ of integrated luminosity.
We demonstrate that in the case of a signal, the higgsino mass can be determined by reconstruction of the dimuon mass edge.
}

\arxivnumber{1211.5584}


\begin{document}

\maketitle

\flushbottom

\section{Introduction}
\label{sec:intro}

The Large Hadron Collider (LHC) at CERN has brought the field of high energy physics into a whole new era.
The observation by both the ATLAS and CMS experiments of a new resonance at $\sim \unit[126]{GeV}$~\cite{:2012gk, :2012gu} is consistent with the long-sought Higgs boson.
If this is indeed the scalar particle needed to induce electroweak symmetry breaking in the standard model (SM), the search for new physics does not end there.
Such a light scalar receives quadratically divergent corrections to its mass, leading to large fine-tuning within the SM.
Furthermore, although the gravitational evidence for dark matter is strong (see \textit{e.g.}~\cite{Bergstrom:2000pn}), its particle nature has yet to be determined.

The minimal supersymmetric standard model (MSSM) addresses both these issues but also faces new problems.
While unbroken supersymmetry only introduces one new parameter, the higgsino mass parameter $\mu$, soft supersymmetry breaking generically introduces many new parameters along with sources of large flavour- and CP-violation (see \textit{e.g.}~\cite{Baer:2006rs}).
The latter problems would be alleviated if the superpartners have large (multi-TeV) masses.
Indeed, this would be consistent with a relatively large value of $\sim \unit[126]{GeV}$ for the Higgs mass and the absence of LHC signals for any other new particles so far.

The disadvantage of such a decoupling solution is that heavy superpartners possibly reintroduce fine-tuning in the electroweak sector.
Nevertheless, a light-higgsino scenario (LHS), in which two higgsino-like neutralinos and a higgsino-like chargino are light (of the order \unit[100]{GeV}) and the other superparticles heavy, is a theoretical possibility.
Experimental limits from chargino searches tell us that $\mu$ cannot be zero and there is \emph{a priori} nothing relating its value to the soft breaking parameters.
Even if the electroweak scale can be obtained with cancelations of terms with large soft parameters, this would be spoiled again if $\mu$ is too large (see~\emph{e.g.}~\cite{Brummer:2012me}).

In fact, a spectrum with higgsino masses around the electroweak scale and the other sparticles typically in the TeV range has been shown recently to arise in models of hybrid gauge-gravity mediated supersymmetry breaking~\cite{Brummer:2011yd}.
Such models are motivated by ways of obtaining the MSSM and grand unification in string theory that feature a hidden sector of exotic states.
They thus make use of the mechanism present in gravity-mediated supersymmetry breaking to generate an electroweak-scale $\mu$, as well as of the advantages of gauge mediation by providing a large number of messengers.

In this set-up we have
\begin{align}
    \abs{ \mu }
  \ll
    \abs{ M_{1,2} }
\ ,
\end{align}
where $M_{1,2}$ are the bino- and wino mass parameters, and mass splittings in the higgsino sector of the order
$\nicefrac{ m_Z^2 }{ \abs{ M_{ 1, 2 } } }$.
Because the higgsinos are nearly mass degenerate and the strongly interacting superparticles are out of reach, such a scenario within the usual MSSM is difficult to probe at the LHC~\cite{Baer:2011ec}.
As will be the subject of this work, the prospects change if we allow for lepton number violating, and therefore R-parity violating, couplings.

R-parity conserves baryon and lepton number and is imposed in the usual MSSM in order to forbid proton decay.
Its conservation also renders the lightest supersymmetric particle stable, making the lightest neutralino a natural WIMP dark matter candidate.
From a theoretical point of view, however, R-parity conservation is not particularly favoured and the stability of the proton can be ensured by requiring either baryon or lepton number violation to be small.
In fact, it is a disadvantage of the MSSM LHS scenario that the relic density of neutralinos is too low, due to coannihilations, to provide the dark matter.
Here, we will consider an extension of the MSSM with broken R-parity where the dark matter candidate is instead the gravitino.

The possibility of gravitino dark matter has an interesting connection to leptogenesis.
While the gravitino is a prediction of the desirable promotion of supersymmetry to a local symmetry, it leads to the gravitino problem~\cite{Weinberg:1982zq,Ellis:1984er,Kawasaki:2004yh,Kawasaki:2004qu,Jedamzik:2006xz}.
The thermal production of gravitinos depends on the reheating temperature and the gravitino and gluino masses~\cite{Bolz:2000fu}.
In thermal leptogenesis, the lepton asymmetry is created through the decays of heavy right-handed neutrinos and then transferred to a baryon asymmetry via sphaleron processes.
In order to generate enough CP asymmetry, as well as to account for the small neutrino masses generated via the seesaw mechanism, the right-handed neutrinos need to be very heavy and therefore high reheating temperatures are required to produce them thermally~\cite{Fukugita:1986hr,Davidson:2002qv,Buchmuller:2004nz,Buchmuller:2005eh}. 
This would also lead to gravitinos being produced in great abundance.
If the gravitino is not the lightest supersymmetric particle (LSP), its
decays would interfere with big bang nucleosynthesis (BBN)~\cite{Kawasaki:1994af}.
If the gravitino is the LSP and a dark matter candidate, the BBN bounds instead apply to the next-to-lightest supersymmetric particle (NLSP) which is meta-stable.
This is where R-parity violation (RPV) comes in.

The requirement of successful baryogenesis puts an upper bound on the amount of RPV that can be allowed, by the condition that the baryon asymmetry is not erased before the electroweak phase transition sets in.
A small amount of RPV, however, leads to 1) a gravitino lifetime exceeding the age of the universe, because of the double suppression of gravitino decays by the Planck mass and the small RPV coupling and 2) a sufficiently short lifetime of the NLSP to be consistent with primordial nucleosynthesis.
This makes it possible to have a good gravitino dark matter candidate even with the high reheating temperatures needed for leptogenesis, thus solving the gravitino problem~\cite{Buchmuller:2007ui}.

It has been shown that the gravitino can account for the observed dark matter abundance for typical gluino masses and different types of NLSPs~\cite{Buchmuller:2008vw}.
In the LHS, we will have to allow for a very large gluino mass and we discuss in this work how the gravitino dark matter abundance can be accounted for in this case.
There are also direct bounds on the RPV couplings from cosmology, which will be of particular relevance for the present work.
Decaying dark matter of this kind would lead to a diffuse gamma-ray flux observable by the Fermi-LAT.
The non-observation of such an excess gives an upper bound on the RPV and thereby a lower bound on the NLSP decay length~\cite{Bobrovskyi:2010ps}.
The finite NLSP decay length leads to the prediction of the displaced vertex signatures at the LHC that we study here.

Displaced vertex signatures are extremely powerful in suppressing SM backgrounds and have been studied, in the context of different SUSY models, for macroscopic decay lengths ranging from \unit[1]{mm} to hundreds of meters~\cite{Ishiwata:2008tp,Asai:2009ka,Meade:2010ji,Bobrovskyi:2011vx,Hirsch:2005ag, Ghosh:2012pq}.
Unlike {\it e.g.}~models where neutrino masses are generated by RPV, which lead to decay lengths up to \unit[1]{mm}, the scenario described above predicts decay lengths that are orders of magnitude larger.
This will lead to displacements of decay vertices in the outer layers of the multipurpose LHC detectors ATLAS and CMS, which motivates the muon signature that we consider.

Previous studies of RPV at the LHC have been motivated by the fact that bounds on sparticle masses can be weakened due to RPV, thus providing a possible explanation to why supersymmetry has not been discovered yet even if it is already being produced at the LHC~\cite{Bobrovskyi:2011vx,Graham:2012th}.
Here, we will show that RPV can also allow for detection of weakly coupled new physics that generically is not  probed by the usual LHC searches.
We will discuss why also the signal we consider would hide from the LHC searches performed so far, and show that the search strategy that we propose here could reveal new physics already in the data accumulated during the LHC runs with proton collisions at \unit[8]{TeV} center-of-mass energy.

In this work, we consider the MSSM extended by bilinear R-parity violating couplings.
These violate lepton number and R-parity by the introduction of only a small number of free parameters.
Baryon number is conserved also at loop level, ensuring the stability of the proton.
The bilinear interactions can be rotated under the symmetries of the theory to allow for a description in terms of trilinear Yukawa interactions, which simplifies the phenomenological analysis.
In this framework, we study the case of a higgsino-like neutralino NLSP and the prospects for discovery with the data from proton collisions at a center-of-mass energy of \unit[8]{TeV}.
All the strongly interacting superpartners are assumed to be out of reach, and the higgsinos would be pair-produced via a virtual $Z$ or $W$~boson.
The heavier neutralino and chargino will decay into the NLSP, and the NLSP will travel in the detector before decaying into SM particles, typically a $W$~boson and a charged lepton in the case of a higgsino NLSP.
NLSP decays into a gravitino LSP are, due to suppression by the Planck mass, orders of magnitude less probable.
A clear signature arises from at least one of the two NLSPs decaying inside the detector, giving rise to two opposite-sign muons from a secondary vertex.

In section~\ref{sec:rpv} we set up the framework of bilinear RPV, derive the branching ratios for the higgsino-like neutralino NLSP and the gravitino LSP and discuss the cosmological bounds on the RPV couplings.
In section~\ref{sec:lhc} we describe the LHC signature and analysis tools, and present the result of our detector level study for a few benchmark models.
We conclude in section~\ref{sec:conclusion}.

\section{Decaying dark matter in the light-higgsino scenario}
\label{sec:rpv}

If we abandon the requirement of R-parity, the additional terms in the MSSM superpotential together with the soft terms introduce 99 new free parameters into the model~\cite{Barbier:2004ez}.
By allowing only for the bilinear terms, baryon number is conserved, and the number of new parameters is reduced to 9 (\textit{c.f.}~\cite{Allanach:2003eb, Barbier:2004ez}).
Such a scenario can be realised through the spontaneous breaking of $\text{B}-\text{L}$, the difference between baryon and lepton number~\cite{Buchmuller:2007ui}.

Compared to the case studied in~\cite{Bobrovskyi:2011vx} where the lightest neutralino was assumed to be bino-like, direct production of higgsino-like neutralinos will have larger cross-sections since they are not suppressed by mixing angles~\cite{Baer:2006rs}.
In this section we derive the branching ratios of relevance for our phenomenological study and introduce the relevant parameters for the study of RPV in the LHS.

\subsection{Bilinear R-parity breaking}

In the extension of the MSSM with bilinear R-parity breaking that we consider, mass mixing terms between lepton and Higgs fields appear in the superpotential%
\footnote{Our notation for Higgs and matter superfields, scalars and left-handed fermions reads: $H_u = (H_u, h_u)$, $L_i = (\tilde l_i, l_i)$, $E_i = (\widetilde{ \bar e }, \bar e)$, etc., where $i$ is the family index. },
\begin{align}
    \Delta W
  = \mu_i H_u L_i
\ ,
\label{rpvw1}
\end{align}
as well as in the scalar potential, induced by supersymmetry breaking,
\begin{align}
 - \Delta \mathcal L
 = B_i H_u \tilde l_i
 + m^2_{ i d } \tilde l^\dagger_i H_d
 + \text{h.c.}
\ .
\label{rpvv1}
\end{align}
These mixing terms, together with the R-parity conserving superpotential,
\begin{align}
    W
  = \mu H_u H_d
  + h_{ i j }^u Q_i U_j H_u
  + h_{ i j }^d D_i Q_j H_d
  + h_{ i j }^e L_i E_j H_d
\ ,
\label{w1rp}
\end{align}
the scalar mass terms,
\begin{align}
  - \mathcal L_\text{M}
 =&\ m^2_u H_u^\dagger H_u
  + m^2_d H_d^\dagger H_d
  + \left( B H_u H_d + \text{h.c.} \right)
 \nonumber \\
 &+ \widetilde m^2_{ l i } \tilde l_i^\dagger \tilde l_i
  + \widetilde m^2_{ e i } \widetilde{ \bar e }_i^\dagger \widetilde{ \bar e }_i
  + \widetilde m^2_{ q i } \widetilde q_i^\dagger \widetilde q_i
  + \widetilde m^2_{ u i } \widetilde{ \bar u }_i^\dagger \widetilde{ \bar u }_i
  + \widetilde m^2_{ d i } \widetilde{ \bar d }_i^\dagger \widetilde{ \bar d }_i
\ ,
\label{scalarmass}
\end{align}
and the standard $\mathrm{SU(3) \times SU(2)\times U(1)_Y}$ gauge interactions define the supersymmetric standard model with bilinear R-parity breaking.
Note that the Higgs mass terms $m_u^2$ and $m_d^2$ contain the contributions both from the superpotential~\eqref{w1rp} and the soft supersymmetry breaking terms.
For simplicity, we have assumed flavour diagonal mass matrices in~\eqref{scalarmass}.

As discussed in~\cite{Bobrovskyi:2010ps}, it is convenient to work in a basis of $\mathrm{SU(2)}$ doublets where the mass mixings $\mu_i$, $B_i$ and $m_{id}^2$ in Eqs.~\eqref{rpvw1} and~\eqref{rpvv1} are traded for R-parity breaking Yukawa couplings.
This can be achieved by field redefinitions:
the standard rotation of the superfields $H_d$ and $l_i$,
\begin{align}
   H_d
 = H_d^\prime
 - \epsilon_i L^\prime_i
   \ , &&
   L_i
 = L^\prime_i
 + \epsilon_i H_d^\prime
   \ , &&
   \epsilon_i
 = \frac{\mu_i}{\mu}
\ ,
\end{align}
followed by a non-supersymmetric rotation involving all scalar $\mathrm{SU(2)}$ doublets,
\begin{align}
   H^\prime_d
 = H^{ \prime \prime }_d
 - \epsilon^\prime_i \tilde l^{ \prime \prime }_i
   \ , &&
   \varepsilon H^*_u
 = \varepsilon H^{ \prime * }_u
 - \epsilon^{ \prime \prime }_i \tilde l^{ \prime \prime }_i
   \ , &&
   \tilde l^\prime_i
 = \tilde l^{ \prime \prime }_i
 + \epsilon^\prime_i H^{ \prime \prime }_d
 + \epsilon^{ \prime \prime }_i \varepsilon H^{ \prime * }_u
\ ,
\end{align}
where $\varepsilon$ is the usual $\mathrm{SU(2)}$ matrix, $\varepsilon = i \sigma^2$, and $\epsilon_i^\prime$ as well as $\epsilon_i^{ \prime \prime }$ are functions of $B$, $B_i$, $m_{ i d }^2$, $\widetilde m_{ l i }^2$, $m_u^2$ and $m_d^2$~\cite{Bobrovskyi:2010ps}.
The virtue of these two rotations are the disappearing mixing terms between the Higgs and the leptons doublets as well as vanishing sneutrino VEVs, which allows to calculate all RPV decays with usual Yukawa-like Feynman diagrams.

The R-parity breaking Yukawa terms contain couplings between gauginos, lepton doublets and Higgs doublets.
After electroweak symmetry breaking one obtains new mass mixings between higgsinos, gauginos and leptons,
\begin{align}
    - \Delta \mathcal L_M
\supset
    m^e_{ij} \frac{\zeta_i}{c_\beta} \overline e_j h_d^-
  - m_Z s_w \zeta_i^* \nu_i b
  + m_Z c_w \zeta_i^* \nu_i w^3
  + \sqrt 2 m_Z c_w \zeta_i^* e_i w^+
  + \text{h.c.}
\ ,
\label{mixing}
\end{align}
where we have defined:
\begin{subequations}
\begin{align}
    \zeta_i
 &= \frac{ \epsilon_i^\prime v_d + \epsilon_i^{ \prime \prime } v_u }{ v }
\ , &&&
    v
 &= \sqrt{ v_u^2 + v_d^2 }
\ , &&&
    \frac{ v_u }{ v_d }
 &= \tan{\beta}
  \equiv
    \frac{s_\beta}{c_\beta}
\ , \label{zetabeta}
 \\ m^e_{ i j }
 &= h_{ i j }^e v_d
\ , &&&
    m_Z
 &= \frac{ \sqrt{ g^2 + g^{ \prime 2 } } v }{ \sqrt 2 }
\ , &&&
    s_w
 &= \frac{ g^\prime }{ \sqrt{ g^2 + g^{ \prime 2 } } }
  = \sqrt{ 1 - c_w^2 }
\ .
\end{align}
\end{subequations}
Here $g$, $g^\prime$ and $h_{ i j }^e$ are the SU(2) and the $\mathrm{U(1)_Y}$ gauge couplings and the charged lepton Yukawa couplings, respectively.
Note that one also obtains couplings of the bino and wino to the lepton doublets and the Higgs doublets \cite{Bobrovskyi:2012dt}
\begin{align}
    - \Delta \mathcal L
  = - \frac{ g^\prime }{ \sqrt 2 }
    \left( \epsilon_i^\prime H_d^{ 0 * } \nu_i
         + \epsilon_i^{ \prime \prime } H_u^0 \nu_i \right) b
         +  \frac{ g }{ \sqrt 2 }
             \left( \epsilon_i^\prime H_d^{ 0 * } \nu_i
                  + \epsilon_i^{ \prime \prime } H_u^0 \nu_i \right) w^3
    + \text{h.c.}
\ ,
\end{align}
where we have shown only the couplings to the neutral Higgs states.
The neutral higgsinos, on the other hand, only couple to the charged Higgs.
Introducing the physical Higgs fields in the unitary gauge, and taking only the coupling to the lightest Higgs into account, one obtains:
\begin{align}
  - \Delta \mathcal L
  = - \frac{ 1 }{ 2 } g^\prime
     \kappa_i
    h \nu_i b
    +  \frac{ 1 }{ 2 } g
     \kappa_i
        h \nu_i w^3
    + \text{h.c.}
\ ,
\label{eq:neutralinohiggscoupl}
\end{align}
where
\begin{align}
    \kappa_i
  = \epsilon_i^\prime \sin( - \alpha )
  + \epsilon_i^{ \prime \prime } \cos( \alpha ).
\end{align}
In the Higgs decoupling limit,  $\alpha \simeq \beta - \nicefrac{ \pi }{ 2 }$, we have $\kappa_i \simeq \zeta_i$.
The Higgs decoupling limit is satisfied in the models considered in the present work.

The details of the following derivation are given in Appendix~\ref{App:BR}.
The gaugino and higgsino mass terms together with the mixing terms in Eq.~\eqref{mixing} represent the $7 \times 7$ neutralino mass matrix in the basis of gauginos $b$, $w^3$, higgsinos $h_u^0$, $h_d^0$ and the three gauge eigenstates of the neutrinos $\nu_i$~\eqref{neutralino mixing matrix}, and also the $5 \times 5$ chargino mass matrix of gaugino, higgsino and the gauge eigenstates of the charged leptons~\eqref{chargino mixing matrix}.
Both mass matrices have to be diagonalized by (bi-) unitary transformation matrices $U$ in order to obtain the mass eigenstates of the neutralinos $\chi^0$ and charginos $\chi^\pm$, respectively.
The currents which couple gauge fields to neutralinos and charginos are modified as well by these transformations and then depend on CKM-type matrix elements of neutral $V^{(\chi^0,\nu)}$, charged $V^{(\chi^0,e)}$, which are functions of the transformation matrices $U$.
Furthermore we have derived the coupling to the supercurrents $U^{(\chi^0,\nu)}$, as well as the coupling to the lightest Higgs $\widetilde V^{(\nu, \chi^0)}$.
These R-parity breaking matrix elements are calculated in Appendix~\ref{App:BR} and read
\begin{subequations}
\begin{align}
    V_{ 1 i }^{ ( \chi^0, \nu ) }
 &= \frac{ \zeta_i m_Z^2}{ 2 \sqrt 2 \mu }
    \left(
      \left( \frac{ s_w^2 }{ M_1 } + \frac{ c_w^2 }{ M_2 } \right) ( s_\beta - c_\beta )
    - \frac{ M_{ \tilde \gamma} - \mu }{ ( M_1 - \mu ) ( M_2 - \mu ) } ( s_\beta + c_\beta )
    \right) \notag \\
& \qquad \times
    \left( 1 + \mathcal O \left( s_{ 2 \beta } \frac{ m_Z^2 }{ \widetilde m^2 } \right) \right)
\ , \label{neutrneut} \\
V_{ 1 i }^{ ( \chi^0, e ) }
  &= \frac{ \zeta_i m_Z^2 }{ \sqrt 2 \mu }
    \left(
      2 \frac{ \mu c_w^2 }{ M_2 ( M_2 - \mu ) } ( s_\beta + c_\beta )
    - \frac{ M_{ \tilde \gamma} - \mu }{ ( M_1 - \mu ) ( M_2 - \mu ) } ( s_\beta + c_\beta )
    - 2 \frac{ c_w^2 }{ M_2 } s_\beta
    \right) \notag \\
    & \qquad \times
        \left( 1 + \mathcal O \left( s_{ 2 \beta } \frac{ m_Z^2 }{ \widetilde m^2 }\right) \right)
    \ , \label{neutrcharged} \\
       \widetilde V^{ ( \nu, \chi^0 ) }_{ i 1 }
      &= - \frac{ \zeta_i m_Z }{ \sqrt 2 } \left( \frac{ c_w }{ M_2 - \mu } + \frac{ s_w t_w }{ M_1 - \mu } \right) ( s_\beta + c_\beta )
          \left( 1 + \mathcal O \left( s_{ 2 \beta } \frac{ m_Z^2 }{ \widetilde m^2 } \right) \right) \ , \label{neutrhiggs}\\
    U^{ ( \widetilde \gamma, \nu ) }_i
 &= \zeta_i m_Z s_w c_w \frac{ M_2 - M_1 }{ M_1 M_2 }
 \left( 1 + \mathcal O \left( s_{ 2 \beta } \frac{ m_Z^2 }{ \widetilde m^2 } \right) \right)
\ , \label{gravphoton}
\end{align}
\end{subequations}
where the photino matrix element and the photino mass parameter are defined as
\begin{align}
    U^{ ( \tilde\gamma, \nu ) }_i
  = c_w U^{(b,\nu)}_i
  + s_w U^{(w,\nu)}_i
\ ,
&&  M_{ \tilde \gamma}
  = M_1 c_w^2 + M_2 s_w^2
\ ,
\end{align}
and $\widetilde m$ is the largest out of the supersymmetric mass parameters $M_1$, $M_2$ and $\mu$ in the neutralino (chargino) mass matrix.
Hence the approximated diagonalisation does not depend on the details of the supersymmetric spectrum, but in fact depends only on the ratio between the electroweak scale and the largest supersymmetric parameter.

\subsection{Gravitino and neutralino decays}
\label{sec:neutralino_signatures}

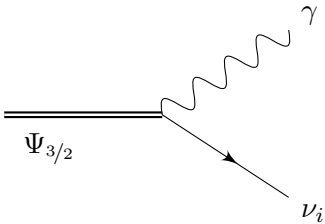
\begin{figure}
\centering
\tikzsetnextfilename{gravitino-decay}
\begin{tikzpicture}[feynman]
\node (gravitino) {} ;
\node (grganu) [right = 1.25*\unitlength of gravitino]{};
\draw [gravitino] (gravitino) node [initial particle name]{$\Psi_{\nicefrac{3}{2}}$}-- (grganu) ;
\node (gamma) [above right = of grganu, final particle name] {$\gamma$};
\draw [boson] (grganu) -- (gamma) ;
\node (nu) [below right = of grganu, final particle name] {$\nu_i$};
\draw [fermion] (grganu) -- (nu) ;
\end{tikzpicture}
\caption{Gravitino decay into photon and neutrino.}
\label{fig:Grav_Photon_Neutr}
\end{figure}

The partial width for a gravitino decaying into a photon and a neutrino (see Figure~\ref{fig:Grav_Photon_Neutr}) is given by~\cite{Takayama:2000uz}
\begin{align}
    \Gamma_\threehalf( \gamma \nu )
  = \frac{ 1 }{ 32 \pi } \sum_i \abs{ U^{ ( \tilde \gamma, \nu ) }_i }^2 \frac{ m_\threehalf^3 }{ M_\text{P}^2 }
\ .
\label{widthgravitino}
\end{align}
Inserting the matrix element \eqref{gravphoton} one obtains for the gravitino lifetime to leading order in $\nicefrac{m_Z}{\tilde m}$~\cite{Bobrovskyi:2010ps}:
\begin{align}
   \Gamma_\threehalf( \gamma \nu )
 = \frac{ 1 }{ 32 \sqrt 2 }
   \frac{ \alpha }{ G_F }
   \zeta^2
   \frac{ m_\threehalf^3 }{ M_\text{P}^2 }
   \left( \frac{ M_2 - M_1 }{ M_1 M_2 } \right)^2
\ ,
\label{Gravitino decay}
\end{align}
where $G_F$ and $\alpha$ are the Fermi constant and the fine structure constant, respectively, and we have introduced the overall R-parity breaking parameter $\zeta$
\begin{align}
    \zeta^2
  = \sum_i \zeta_i^2
\ .
\end{align}
Contrary to the bino-like neutralino case (\textit{c.f.}~\cite{Bobrovskyi:2010ps, Bobrovskyi:2011vx}) the gravitino lifetime cannot be expressed directly in terms of the mass of the  lightest neutralino.
Instead it depends on the mass scale of the gaugino mass parameter and hence on the masses of the heavier neutralinos.
Nevertheless, one can invert the relation \eqref{Gravitino decay} with respect to the R-parity breaking parameter
\begin{align}
    \zeta
  = 4 \times 2^{3/4}
    \sqrt \frac{ G_F }{ \alpha }
    \frac{ M_\text{P} }{ \sqrt{ m_\threehalf^3 \tau_\threehalf } }
    \frac{ M_1 M_2 }{ M_2 - M_1 }
\ ,
\label{zetafromgrav}
\end{align}
and constrain its value from the bounds on the gravitino lifetime using reasonable assumptions for the bino and wino mass parameters (see Section~\ref{sec:fermibounds}).

\begin{figure}
\centering
\tikzsetnextfilename{neutralino-decay-Z}
\begin{tikzpicture}[feynman]
\node (neutralino) {} ;
\node (grZnu) [right = 1.25*\unitlength of neutralino]{};
\draw [anti neutralino] (grZnu) -- (neutralino)  node [initial particle name]{$\chi_1^0$};
\node (Z) [above right = of grZnu, final particle name] {$Z$};
\draw [boson] (grZnu) -- (Z) ;
\node (nu) [below right = of grZnu, final particle name] {$\nu_i$};
\draw [fermion] (grZnu) -- (nu) ;
\end{tikzpicture}
\hfill
\tikzsetnextfilename{neutralino-decay-W}
\begin{tikzpicture}[feynman]
\node (neutralino) {} ;
\node (grWnu) [right = 1.25*\unitlength of neutralino]{};
\draw [anti neutralino] (grWnu)-- (neutralino)  node [initial particle name]{$\chi_1^0$};
\node (W) [above right = of grWnu, final particle name] {$W^\pm$};
\draw [boson] (grWnu) -- (W) ;
\node (nu) [below right = of grWnu, final particle name] {$l_i^\mp$};
\draw [fermion] (grWnu) -- (nu) ;
\end{tikzpicture}
\hfill
\tikzsetnextfilename{neutralino-decay-h}
\begin{tikzpicture}[feynman]
\node (neutralino) {} ;
\node (grhnu) [right = 1.25*\unitlength of neutralino]{};
\draw [anti neutralino] (grhnu)-- (neutralino)  node [initial particle name]{$\chi_1^0$};
\node (h) [above right = of grhnu, final particle name] {$h$};
\draw [scalar] (grhnu) -- (h) ;
\node (nu) [below right = of grhnu, final particle name] {$\nu_i$};
\draw [fermion] (grhnu) -- (nu) ;
\end{tikzpicture}
\caption{Neutralino decays into neutrino and $Z$~boson, charged lepton and $W$~boson, and neutrino and the lightest Higgs boson.}
\label{fig:Neutralino_Decays}
\end{figure}
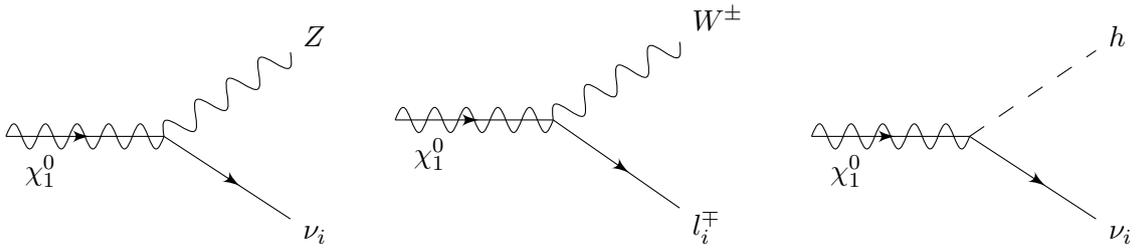

A neutralino with a mass larger than \unit[100]{GeV} decays predominantly via two-body channels either into a $W$~boson and a charged lepton, a $Z$~boson and a neutrino~\cite{Mukhopadhyaya:1998xj} or a neutrino and a Higgs boson (see Figure~\ref{fig:Neutralino_Decays}).
In the case of the higgsino-like neutralino the decay into a $W$~boson and a charged lepton is dominant.
For completeness we present the results for all three channels.
The partial decay widths%
\footnote{Decays into antiparticles have been taken into account.}%
read
\begin{subequations}
\begin{align}
   \Gamma \left( \chi_1^0 \rightarrow W^\pm l^\mp \right)
=& \frac{ G_F }{ 4 \sqrt 2 \pi } m_{ \chi_1^0 }^3 \sum_i
   \abs{ V_{ 1 i }^{ ( \chi^0, e ) } }^2
   \left(1 - \frac{ m_W^2 }{ m_{ \chi^0_1 }^2 } \right)^2
       \left( 1 + 2 \frac{ m_W^2 }{ m_{ \chi^0_1 }^2 } \right)
\ ,
\\ \Gamma \left( \chi_1^0 \rightarrow Z \nu \right)
=& \frac{ G_F }{ 2 \sqrt 2 \pi } m_{ \chi_1^0 }^3 \sum_i
   \abs{ V_{ 1 i }^{ ( \chi^0, \nu ) } }^2
   \left( 1 - \frac{ m_Z^2 }{ m_{ \chi^0_1 }^2 } \right)^2
       \left( 1 + 2 \frac{ m_Z^2}{ m_{ \chi^0_1 }^2 } \right)
\ ,
\\    \Gamma \left( \chi_1^0 \rightarrow h \nu \right)
    =& \frac{ \alpha  }{ 16 s_w^2 } m_{ \chi_1^0 } \sum_i
       \abs{ \widetilde V_{ 1 i }^{ ( \nu, \chi^0 ) } }^2
       \left( 1 - \frac{ m_h^2}{ m_{ \chi^0_1 }^2} \right)^2 \ ;
\end{align}
\label{Neutralino_Decay_Widths}
\end{subequations}
where $V_{1i}^{(\chi^0,e)}$ and  $V_{1i}^{(\chi^0,\nu)}$ are the charged and neutral current matrix elements, whereas $\widetilde V_{1i}^{(\nu, \chi^0)} $ is the matrix element for the decay into the Higgs boson.
We will evaluate these elements at the leading order using the expressions given in Eqs.~\eqref{neutrneut}, \eqref{neutrcharged} and \eqref{neutrhiggs}, respectively. We will assume throughout the rest of this work that the Higgs decoupling limit is satisfied.

\begin{figure}
\renewcommand{\Width}{.375\textwidth}
\small
\hfill
\begin{xsubcaption}[.48\textwidth]
\tikzsetnextfilename{Higgsino-Phasespace-factors}
\newcommand{\xMax}{ 600 } 


\newcommand{\mW}{ 80.4 } 
\newcommand{\mZ}{ 91.2 } 
\newcommand{\mH}{ 125 } 


\newcommand{\Ratio}[1]{ ( #1^2 / x^2 ) } 
\newcommand{\PsS}[1]{ ( 1 - \Ratio{ #1 } )^2 } 
\newcommand{\PsV}[1]{ ( \PsS{ #1 } * ( 2 * \Ratio{ #1 } + 1 )  ) } 

\begin{tikzpicture}
\begin{axis}[ 
  xmin = \mW
, xmax = \xMax
, xlabel = { Neutralino mass [GeV] }
, ymin = 0, ymax = 1
, legend pos = south east
]
\addplot [ domain = \mW : \xMax, plotone ]{ \PsV{ \mW } } ;
\addlegendentry{$W^\pm$}
\addplot [ domain = \mZ : \xMax, plottwo ]{ \PsV{ \mZ } } ;
\addlegendentry{$Z$}
\addplot [ domain = \mH : \xMax, plotthree ]{ \PsS{ \mH } } ;
\addlegendentry{$h$}
\end{axis}
\end{tikzpicture}
\caption{Phase space suppression factors for neutralino decays.}
\label{fig:Higgsino-Phasespace}
\end{xsubcaption}
\hfill
\begin{xsubcaption}[.48\textwidth]
\tikzsetnextfilename{Higgsino-Branching-Ratio}
\newcommand{\xMax}{ 600 } 
\newcommand{\LogSave}{ .001 } 


\newcommand{\mW}{ 80.4 } 
\newcommand{\mZ}{ 91.2 } 
\newcommand{\mH}{ 125 } 
\newcommand{\mBino}{ 3000 } 
\newcommand{ \betavalue }{ ( atan( 50 ) ) } 


\newcommand{\Ratio}[1]{ ( #1^2 / x^2 ) } 
\newcommand{\PsS}[1]{ ( 1 - \Ratio{ #1 } )^2 } 
\newcommand{\PsV}[1]{ ( \PsS{ #1 } * ( 2 * \Ratio{ #1 } + 1 )  ) } 
\newcommand{\DwBnA}{ ( 2 * \PsV{ \mW}  + \PsV{ \mZ } ) } 
\newcommand{\DwBnB}{ ( \DwBnA + \PsS{ \mH } ) } 
\newcommand{\DwHnA}{ ( 81 * \PsV{ \mW } ) } 
\newcommand{\DwHnB}{ ( \DwHnA + 8 * \PsS{ \mH } ) } 
\newcommand{\ZSupr}{  ( x^2 * ( 17 - 8 * .5376 )^2 ) / ( 162 * \mBino^2 * .591 )  } 

\newcommand{\mA}{ ( 159 + 3.42 * x ) } 
\newcommand{\alphavalue}{ ( ( \mA^2 + \mZ^2) / ( \mA^2 - \mZ^2 ) * tan( 2 * \betavalue ) ) } 

\begin{tikzpicture}
\begin{semilogyaxis}[ 
  xmin = \mZ
, xmax = \xMax
, xlabel = { Neutralino mass [GeV] }
, ymin = .000001
, ymax = 2
, try min ticks log = 5
, legend pos = south east
]
\addplot [ domain = \mZ : \mH, plotone ]{ 1 } ;
\addplot [ domain = \mH : \xMax, plotone, forget plot ]{ 81 * \PsV{ \mW } / \DwHnB } ;
\addlegendentry{$W^\pm$}
\addplot [ domain = \mZ + \LogSave : \mH, plottwo ]{ \ZSupr * \PsV{ \mZ } / \DwHnA } ;
\addplot [ domain = \mH : \xMax, plottwo, forget plot ]{ \ZSupr * \PsV{ \mZ } / \DwHnB } ;
\addlegendentry{$Z$}
\addplot [ domain = \mH + \LogSave : \xMax, plotthree ]{ 8 * \PsS{ \mH } / \DwHnB } ;
\addlegendentry{$h$}
\end{semilogyaxis}
\end{tikzpicture}
\caption{Neutralino branching ratios.}
\label{fig:Higgsino-Branchingratio}
\end{xsubcaption}
\hfill\strut
\caption{Phase space  suppression factors and branching ratios for neutralino decays into $W$~bosons (black), $Z$~bosons (red) and the lightest Higgs (blue) and leptons.}
\label{fig:Higgsino-Plots}
\end{figure}

Figure~\ref{fig:Higgsino-Phasespace} shows the phase space suppression factors that are important for small neutralino masses.
We have assumed that the mass of the lightest Higgs boson is \unit[125]{GeV}.
The total neutralino decay width is given by the sum
\begin{align}
   \Gamma_{ \chi_1^0 }
 = \Gamma( \chi_1^0 \rightarrow W^\pm l^\mp )
 + \Gamma( \chi_1^0 \rightarrow Z \nu )
 + \Gamma( \chi_1^0 \rightarrow h \nu )
\ .
\label{Neutralino_Decay_Width}
\end{align}
However, the evaluation of the partial decay widths in the case of a higgsino-like neutralino (see Figure~\ref{fig:Higgsino-Branchingratio}) leads to the conclusion that the lifetime of the lightest higgsino can be estimated solely from the decay into a $W$~boson and a charged lepton.
Below, we only provide the approximative formula, where we have taken into account that $\tan \beta$ in hybrid gauge-gravity mediation is in general large.
In the numerical evaluation, however, we use the full results.
Using Eq.~\eqref{zetafromgrav} we can express $\zeta$ in terms of the gravitino lifetime and arrive at the following expression for the higgsino-like neutralino lifetime
\begin{align}
    \tau_{ \chi^0_1 }
  \simeq&\
    \frac{ 1 }{ 2 \pi }
    \frac{ m_\threehalf^3 \tau_\threehalf }{ M_\text{P}^2 }
    \frac{ c_w^4 s_w^4 }{ \alpha }
    \frac{ 1 }{ f_W( m_{ \chi^0_1 } ) }
    \frac{ \left ( M_2 - M_1 \right )^2 }{ M_1^2 M_2^2 \mu } \\
 &\times
    \left(
      2 \frac{ \mu c_w^2 }{ M_2 ( M_2 - \mu ) } ( s_\beta + c_\beta )
    - \frac{ M_{ \tilde \gamma} - \mu }{ ( M_1 - \mu ) ( M_2 - \mu ) } ( s_\beta + c_\beta )
    - 2 \frac{ c_w^2 }{ M_2 } s_\beta
    \right)^{ -2 }
\notag \ ,
\end{align}
where $f_W$ is the phase space suppression factor appearing in Eq.~\eqref{Neutralino_Decay_Widths}
\begin{align}
    f_W( m_{ \chi_1^0 } )
  = \left( 1 - \frac{ m_W^2 }{ m_{ \chi_1^0 }^2 } \right)^2
    \left( 1 + 2 \frac{ m_W^2 }{ m_{ \chi_1^0 }^2 } \right)
\ ,
\end{align}
and we have set $m_{\chi^0_1} \simeq \mu$.
The neutralino lifetime depends on the neutralino mass, the gravitino mass and its lifetime, and additionally on the bino and wino mass parameters.
Expanding the higgsino lifetime in $\nicefrac{ \mu }{ \tilde m }$ allows us to arrive at a formula which is to first order independent of the higher neutralino mass scale $\widetilde m \sim M_1 \sim M_2$:
\begin{align}
    \tau_{ \chi^0_1 }
  \simeq&\
    \frac{ 1 }{ 2 \pi }
    \frac{ m_{ \nicefrac{ 3 }{ 2 } }^3 \tau_{ \nicefrac{ 3 }{ 2 } } }{ M_\text{P}^2 }
    \frac{ s_w^4 c_w^4 }{ \alpha }
    \frac{ 1 }{ f_W( m_{ \chi^0_1 } ) }
    \frac{ 1 }{ m_{ \chi^0_1 } }
    \left(
      \frac{ M_2 - M_1 }{ 3 M_1 c_w^2 + M_2 s_w^2 }
    \right)^2
    \left( 1
    + \mathcal O \left( \frac{ m_{ \chi^0_1 } }{ m_{ \chi^0_4 } } \right)
    \right)
\ .
\label{neutralino lifetime}
\end{align}
The factor including the gaugino masses $M_1$ and $M_2$ depends only on their ratio and is \textit{e.g.}~in the case of unified GUT masses very close to $\nicefrac{ 1 }{ 8 }$.

\subsection{Cosmological bounds}

Both lower and upper bounds on $\zeta$ can be derived from cosmology.
The lower bound comes from the BBN contraints on the NLSP when the gravitino is the LSP.
An upper bound can in principle be derived by demanding that the baryon asymmetry generated by leptogenesis is not washed out before the electroweak phase transition in the early universe~\cite{Campbell:1990fa,Fischler:1990gn,Dreiner:1992vm,Endo:2009cv}.
However, the bound from the constraints on decaying dark matter from the Fermi gamma-ray searches is stronger.
As we will see, for our analysis the lower bound is not very constraining while the upper bound will be the motivation for our LHC search strategy.

Having derived the decay widths of the gravitino LSP and the higgsino NLSP, we are now ready to estimate the gravitino mass range allowing for gravitino dark matter and successful leptogenesis~\cite{Buchmuller:2008vw}.
This allows us to connect the results from gamma-ray searches with displaced neutralino decays at the LHC.

\subsubsection{Big bang nucleosynthesis}

To start with, we need to make sure that the decays of higgsino NLSPs do not interfere with BBN.
Hence we demand that all higgsinos decay during the first 100 seconds of the universe~\cite{Feng:2004zu, Kawasaki:2004qu, Steffen:2006hw}.
Deriving the neutralino lifetime \eqref{neutralino lifetime} without substituting the R-parity violating parameter $\zeta$ for the gravitino mass and lifetime leads to
\begin{align}
    \zeta
  \simeq
    4.23 \times 10^{-12}
    \left( \frac{ \tau_{\chi_1^0} }{ \unit[100]{s} } \right)^{-\onehalf}
    \left( \frac{ m_{ \chi_1^0 } }{ \unit[100]{GeV} } \right)^{-\onehalf}
    \left( \frac{ m_{ \chi_3^0 } }{ \unit[2]{TeV} } \right)
\ ,
\label{zeta from neutralino properties}
\end{align}
which then characterizes the lower bound on $\zeta$.

\subsubsection{Gravitino dark matter mass}

The minimal gravitino mass is limited by the requirement that the gravitino abundance does not overclose the universe.
Since gravitinos are produced in thermal SQCD scatterings $gg \rightarrow \tilde g \Psi_\mu$~\cite{Bolz:2000xi}, the gravitino mass must increase with increasing gluino mass for a given reheating temperature.
As we are interested in models in which the colored particles are inaccessible at the LHC, gluinos will typically be very heavy.
For example, in the hybrid gauge-gravity mediation scenario in~\cite{Brummer:2012zc}, which gives rise to a Higgs mass close to the tentative LHC result, the gluino mass is close to \unit[4]{TeV}.
In order to still allow for small gravitino masses, we will assume that the hot phase of the universe was created in the decay of the false vacuum of unbroken $\text{B}-\text{L}$~\cite{Buchmuller:2011mw, Buchmuller:2012wn}.
Since right-handed neutrinos are created from $\text{B}-\text{L}$ Higgs decays, this scenario allows for gravitino dark matter, leptogenesis and the correct values for the neutrino mass parameters while requiring lower reheating temperatures compared to the thermal leptogenesis case.
The lower bound on the gravitino mass obtained in~\cite{Buchmuller:2012wn} for $m_{ \tilde g } = \unit[1]{TeV}$ is $m_\threehalf^\text{min} = \unit[10]{GeV}$.
It is possible to scale this bound to other gluino masses using~\cite{Buchmuller:2011mw}
\begin{align}
    m_\threehalf
  = m_\threehalf^\text{min} \left( \frac{ m_{ \tilde g } }{ \unit[1]{TeV} } \right)^2
\ .
\end{align}
Assuming a lower bound on the gluino mass of \unit[2]{TeV}, the minimal gravitino mass is \unit[40]{GeV} and therefore a neutralino NLSP with a mass of \unit[100]{GeV} is viable.

\subsubsection{Fermi-LAT bound on the gravitino lifetime}
\label{sec:fermibounds}

With the help of Fermi-LAT data we are able to restrict the lifetime of gravitinos for a given mass.
Using the isotropic diffuse gamma-ray flux one can derive a lower bound of $\tau_\threehalf \gtrsim \unit[3 \times 10^{28}]{s}$~\cite{Bobrovskyi:2010ps}.
A stronger bound of $\tau_\threehalf \gtrsim \unit[6 \times 10^{28}]{s}$ can be derived as a consequence of the non-observation of any gamma-ray lines~\cite{Vertongen:2011mu, Ackermann:2012qk, Cirelli:2012ut}.
For a LHS with a bino mass of roughly $M_1 \sim \unit[2]{TeV}$ this translates via \eqref{zetafromgrav} into an upper bound on the R-parity violation of $\zeta \lesssim 4.70 \times 10^{-8}$ and $\zeta \lesssim 3.32 \times 10^{-8}$, respectively.%
\footnote{This stringent RPV bound implies that the neutrino masses are dominated by the contribution from right handed neutrinos \cite{Buchmuller:2007ui}.}%
Finally we are able to derive a lower bound on the decay length of the lightest higgsino as a function of its mass as well as the mass and lifetime of the gravitino
\begin{align}
    c \tau_{ \chi^0_1 }
\gtrsim
    \unit[6.5]{m}
    \left( \frac{ m_{ \chi^0_1 } }{ \unit[400]{GeV} } \right)^{-1}
    \left( \frac{ m_\threehalf }{ \unit[40]{GeV} } \right)^3
    \left( \frac{ \tau_\threehalf ( \gamma \nu ) }{ \unit[10^{28}]{s} } \right)
    f_W( m_{ \chi^0_1 } )^{ -1 }
\ ,
\label{higgsino lieftime}
\end{align}
which is well within the reach of the multipurpose detectors at the LHC.
Even when the higgsino decay length is larger than the detector dimensions, some higgsinos would, due to the statistical nature of the process, decay inside the detector.

\section{Higgsino production and decay at the LHC}
\label{sec:lhc}

In the LHS, higgsino-like charginos and neutralinos would be pair produced at the LHC via virtual $Z$ and $W$~bosons.
Heavier higgsinos decay into lighter ones, the lightest one being the neutralino NLSP.
The mass difference between the lightest chargino and the lightest neutralino is
\begin{align}
    m_{\chi_1^\pm} - m_{\chi_1^0}
  \stackrel{m_Z \ll \tilde m}{\simeq} &\
    \frac{1}{2} m_Z^2 \left( \frac{ ( 1 + s_{2\beta} ) ( M_{ \tilde \gamma} - \mu ) }{ ( M_1 - \mu ) ( M_2 - \mu ) } -  \frac{ 2 c_w^2 ( M_2 s_{ 2 \beta } + \mu ) }{ ( M_2 + \mu ) ( M_2 - \mu ) } \right) \notag \\
 \stackrel{\mu \ll \tilde m}{\simeq} & \
    \frac{1}{2} m_Z^2 \left( \frac{ M_1 c_1 + M_2 c_2 }{ M_1 M_2 }
   + \mu \frac{ M_1^2 c_1 + M_2^2 c_2 }{ M_1^2 M_2^2 } + \mathcal O \left( \frac{ \mu }{ \widetilde m } \right)^2 \right)
\label{chargino mass difference}
\ ,
\end{align}
where we have used
\begin{align}
    c_1
  = ( 1 - s_{2 \beta} ) c_w^2
\ ,
 && c_2
  = ( 1 + s_{2\beta} ) s_w^2
\ .
\end{align}
The mass difference between the next-to-lightest neutralino and the lightest neutralino is
\begin{align}
    m_{\chi_2^0} - m_{\chi_1^0}
  \stackrel{m_Z \ll \tilde m}{\simeq} & \
    \frac{1}{2} m_Z^2 \left( \frac{ ( 1 + s_{2\beta} ) ( M_{ \tilde \gamma} - \mu ) }{ (M_1 - \mu) ( M_2 - \mu ) } + \frac{ ( 1 - s_{2\beta} ) ( M_{ \tilde \gamma} + \mu ) }{ ( M_1 + \mu) ( M_2 + \mu ) } \right) \notag \\
  \stackrel{\mu \ll \tilde m}{\simeq} & \
    m_Z^2 \frac{ M_{\tilde\gamma} }{ M_1 M_2 }
   \left( 1 + \mu s_{ 2 \beta } \left( \frac{ 1 }{ M_1 } + \frac{ 1 }{ M_2 } - \frac{ 1 }{ M_{ \tilde \gamma } } \right)
   + \mathcal O \left(\frac{ \mu }{ \widetilde m } \right)^2 \right)
\label{neutralino mass difference}
\ .
\end{align}
Hence, in the case of heavy gauginos the mass difference is rather small and to first order proportional to $\nicefrac{m_Z}{\tilde m}$.
Therefore standard model products at this stage of the decay chain will be too soft to be detectable.
In the presence of RPV, however, the NLSP would travel in the detector and further decay, prominently into a $W$~boson and a lepton, yielding detectable SM objects coming from a displaced vertex.
The lifetime of the lightest higgsino \eqref{higgsino lieftime} is as a function of $\zeta$
\begin{align}
    c \tau_{ \chi^0_1 }
\gtrsim
    \unit[4.3]{m}
    \left( \frac{ \zeta }{ 10^{ -7 } } \right)^{-2}
    \left( \frac{ m_{ \chi^0_1 } }{\unit[400]{GeV}} \right)^{-1}
    \left( \frac{ m_{ \chi^0_3 } }{\unit[2]{TeV}} \right)^2
    f_W( m_{ \chi^0_1 } )^{ -1 }
\ .
\end{align}

\subsection{Signatures and search strategy}\label{sec: search strategy}

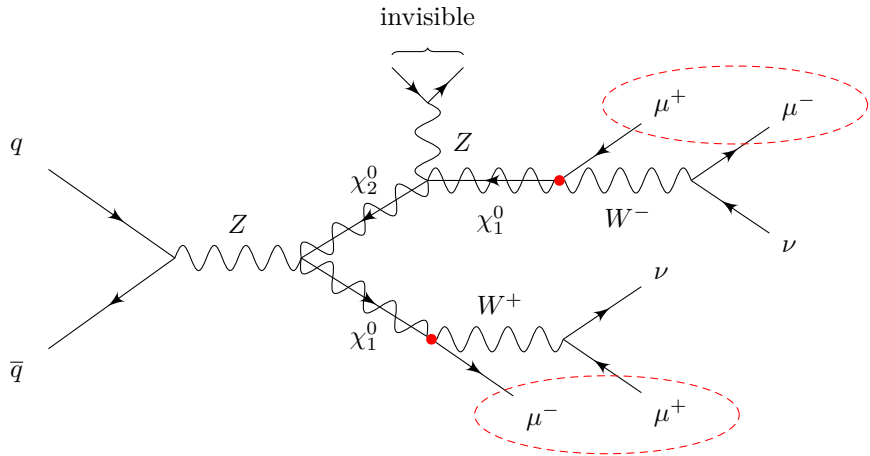
\begin{figure}
\centering
\small
\tikzsetnextfilename{higgsino-rpv-decaychain}
\begin{tikzpicture}[feynman]
\node (q1) [ particle name ] {$q$} ;
\node (qqZ) [ below right = of q1 ] {} ;
\draw [fermion] (q1) -- (qqZ) ;
\node (q2) [particle name, below left= of qqZ ] {$\overline q$} ;
\draw [anti fermion] (q2) -- (qqZ) ;
\node (Zn1n2) [right = of qqZ] {};
\draw [boson] (qqZ) -- node [particle name, above] {$Z$} (Zn1n2) ;
\node (n2Zn1) [above right = of Zn1n2 ] {} ;
\draw [anti neutralino] (Zn1n2) -- node [particle name, above ] {$\chi_2^0$} (n2Zn1) ;
\node (Znunu) [above = of n2Zn1] {} ;
\draw [boson] (n2Zn1) -- node [particle name, right] {$Z$} (Znunu) ;
\node (nu3) [ final particle name, above left = .4\unitlength of Znunu] {};
\draw [anti fermion] (Znunu) -- (nu3) ;
\node (nu4) [ final particle name, above right = .4\unitlength of Znunu]{};
\draw [ decoration = { brace }, decorate ] (nu3) -- node [ anchor = north, yshift= \smaller*\smaller*\unitlength ] {invisible} (nu4);
\draw [fermion] (Znunu) -- (nu4) ;
\node (n1Wmu1) [interaction, right = of n2Zn1] {};
\draw [anti neutralino] (n2Zn1) -- node [particle name, below ] {$\chi_1^0$} (n1Wmu1) ;
\node (mu1) [final particle name,   above right = of n1Wmu1] {$\mu^+$};
\draw [anti fermion] (n1Wmu1) -- (mu1) ;
\node (Wmunu) [ right = of n1Wmu1] {};
\draw [boson, decoration = { mirror }] (n1Wmu1) -- node [particle name, below ] {$W^-$} (Wmunu) ;
\node (mu2) [final  particle name,   above right = of Wmunu] {$\mu^-$};
\draw [fermion] (Wmunu) -- (mu2) ;
\node (nu1) [final  particle name,  below right = of Wmunu] {$\nu$};
\draw [anti fermion] (Wmunu) -- (nu1) ;
\node (n1Wmu2) [ interaction, below right = of Zn1n2] {};
\draw [neutralino, decoration = { mirror }] (Zn1n2) -- node [particle name, below ] {$\chi_1^0$} (n1Wmu2) ;
\node (mu3) [ final particle name, below right = of n1Wmu2] {$\mu^-$};
\draw [fermion] (n1Wmu2) -- (mu3) ;
\node (Wmunu2) [ right = of n1Wmu2] {};
\draw [boson] (n1Wmu2) -- node [particle name, above ] {$W^+$} (Wmunu2) ;
\node (mu4) [final particle name, below right = of Wmunu2] {$\mu^+$};
\draw [anti fermion] (Wmunu2) -- (mu4) ;
\node (nu2) [final particle name, above right = of Wmunu2] {$\nu$};
\draw [fermion] (Wmunu2) -- (nu2) ;
\node [plottwo, draw, ellipse, fit = (mu1) (mu2) ] {};
\node [plottwo, draw, ellipse, fit = (mu3) (mu4) ] {};
\end{tikzpicture}
\caption{Typical R-parity violating decay chain involving higgsino-like neutralinos at the LHC.
The secondary vertices as well as the two possibilities of interesting muon combinations are highlighted.
The $Z$~boson decay is invisible, due to the small mass difference between the heavier higgsinos and the lightest higgsino (see Eqs.~\eqref{chargino mass difference}, \eqref{neutralino mass difference} and Table~\ref{tab:higgsino masses}).
The signature is essentially the same for chargino production, since also in this case the decays into the lightest higgsino lead only to particles with small $p_T$.%
}
\label{fig:Higgsino decay chain}
\end{figure}

As shown in section~\ref{sec:rpv}, neutralinos that are produced in proton collisions at the LHC decay in a secondary vertex into a $W$~boson and a charged lepton in almost \unit[100]{\%} of all decays.
Figure~\ref{fig:Higgsino decay chain} shows an example of a cascade decay with muons in the final state.
The distance between the collision point and the secondary vertex depends on the decay width of the neutralino~\eqref{Neutralino_Decay_Width}, and hence on the R-parity breaking parameter~$\zeta$.

The larger the value of $\zeta$, the larger is the probability that the NLSP decays inside the detector. There are two multipurpose experiments at the LHC, the ATLAS and CMS detectors. Each detector consists of several subdetectors, from the inner detector, for track reconstruction and primary and secondary vertex reconstruction, to the calorimeters and outermost the muon system.
Since the models under study here give rise to rather large displacements, we will choose to rely on the identification of muon objects in our search strategy.
ATLAS has larger dimensions than CMS, with its muon system stretching out to a radius of about \unit[20]{m}~\cite{Aad:2008zzm}.
In our analysis, we choose to use requirements on the radial and beam-line coordinates corresponding to the CMS detector geometry, which will then be the somewhat more conservative choice.
We do not expect our results to depend much on which of the two detectors is considered.
The different detector layers of the CMS detector are~\cite{Bayatian:2006zz}:
\begin{itemize}
\item The inner detector or tracker, 
which stretches out to a radius of $r \sim \unit[110]{cm}$ transverse to the beam. Its innermost part, the pixel detector, covers $r \leq \unit[11]{cm}$.
\item The electromagnetic calorimeter which measures electron and photon energies and stretches to $ r \lesssim \unit[2]{m}$.
\item The hadronic calorimeter, for measuring strongly interacting particles and identification of jets, which stretches out to $r \lesssim \unit[3]{m}$.
\item The magnet, stretching out to $r \sim \unit[4]{m}$.
\item The system of muon detectors or muon chambers, for identification of muons and measurement of their momentum, out to radius of $r \sim \unit[7.4]{m}$.
\end{itemize}

\begin{table}
\centering
\begin{tabular}{l r@{ $\rightarrow$ }l r}
   \toprule
   category
 & \multicolumn{2}{ c }{$\chi^0_1$ decays}
 & LHC signature
\\ \midrule
   leptonic
 & $ W^+ W^-  l^+ l^- $
 & $ l^+\nu l^-\bar \nu l^+ l^-  $
 & \multirow{3}{*}{$2 l^- + 2 l^+ + \slashed E_T $}
\\
 & $ W^+ W^+  l^-l^- $
 & $ l^+\nu l^+\nu l^- l^-  $
 &
\\
 & $ W^- W^-  l^+ l^+ $
 & $ l^-\bar\nu l^-\bar\nu l^+ l^+ $
 &
\\ \midrule
   semi-leptonic
 & $ W^+ W^-  l^+ l^- $
 & $ j j l^-\bar\nu l^+ l^- $
 & \multirow{2}{*}{$ 2 j + 2 l^- + 1 l^+ + \slashed E_T $}
\\
 & $ W^+ W^+  l^- l^- $
 & $ j j l^+\nu  l^- l^- $
 &
\\ \cmidrule{2-4}
 & $ W^+ W^-  l^+ l^- $
 & $ j j l^-\bar\nu l^+l^- $
 & \multirow{2}{*}{$2 j + 1 l^- + 2 l^+ + \slashed E_T $}
\\
 & $ W^- W^- l^+ l^+ $
 & $ j j l^-\bar\nu l^+ l^+ $
 &
\\ \cmidrule{2-4}
 & $ W^+ W^- l^- l^+ $
 & $ j j j j l^- l^+ $
 & $4 j + 1l^- + 1l^+ \phantom{ \; + \slashed E_T } $
\\ \cmidrule{2-4}
  \multirow{2}{*}{\emph{(same sign, no $\slashed E_T $)}}
 & $ W^+ W^+  l^- l^- $
 & $ j j j j l^- l^- $
 & $4 j + 2 l^- \phantom{ \; + \slashed E_T } $
\\
 & $ W^- W^- l^+ l^+ $
 & $ j j j j l^+ l^+ $
 & $4 j + 2 l^+ \phantom{ \; + \slashed E_T } $
\\ \bottomrule
\end{tabular}
\caption{All possible final states in the higgsino-like neutralino case if both neutralinos decay inside the tracking volume.}
\label{tab:finstateshiggsino}
\end{table}

Table~\ref{tab:finstateshiggsino} summarizes all possible LHC signatures if the NLSP is a higgsino-like neutralino in the case when both neutralinos decay inside of the tracker volume.
The signatures are classified according to the final states in the neutralino decays. Unlike the previously studied case of bilinear RPV in the MSSM in~\cite{Bobrovskyi:2011vx}, when the spectra contain light squarks and gluinos and a bino-like neutralino NLSP, there are only two types of signatures: leptonic signatures involving only charged leptons in the final state, and semi-leptonic signatures involving at least two charged leptons and jets.

Most supersymmetry searches for such final states at the LHC so far rely on the large production cross sections of the strongly interacting squarks and/or gluinos.
The searches that are starting to probe direct electroweak production of charginos and neutralinos have been interpreted in R-parity conserving models with a stable neutralino LSP~\cite{:2012gg, :2012ku, :2012ew}.
They rely on leptonic signatures with larger missing transverse energy than what would be present in our scenario in the case of both neutralinos decaying inside the tracker.
Searches in the first LHC data for RPV have also been performed \cite{ATLAS-CONF-2012-153, :2012vd, Aad:2012zx, ATLAS:2012dp, ATLAS:2012kr, Aad:2012yw, Aad:2011qr, CMS-PAS-SUS-12-027} but because of differences in the scenarios considered and differing signatures they do not apply here.
Searches for long-lived neutral particles have been carried out as well but do not apply to our model as they
assume either the wrong event topologies \cite{ATLAS:2012av,Aad:2012kw}, final states \cite{Chatrchyan:2012ir}
and/or size of the displacements \cite{Aad:2012kw,:2012ae}.

\begin{table}
\centering
\begin{tabular}{l r@{ $\rightarrow$ }l r}
  \toprule
   category
 & \multicolumn{2}{c}{$\chi^0_1$ decays}
 & LHC signature
\\ \midrule
   leptonic
 & $W^+ l^- $\
 & $l^+\nu l^-$
 & \multirow{2}{*}{$1 l^- + 1 l^+ + \slashed E_T $}
\\ \emph{(opposite sign)}
 & $W^- l^+ $
 & $l^- \bar\nu l^+$
 &
\\ \midrule
   single lepton
 & $W^+ l^- $
 & $j j l^- $
 & $2 j + 1 l^- + \slashed E_T $
\\
 & $ W^- l^+ $
 & $ jj l^+ $
 & $ 2j + 1 l^+ + \slashed E_T $
\\ \bottomrule
\end{tabular}
\caption{All possible final states in the higgsino-like neutralino case if one of the neutralinos decays outside the tracking volume.}
\label{tab:finstatesouthiggsino}
\end{table}
\begin{figure}
\small
\renewcommand{\Width}{.32*\textwidth}
\renewcommand{\Height}{\Width}
\hfill
\begin{xsubcaption}
\tikzsetnextfilename{misspt}
\newcommand{\crosssection}{7649} 
\newcommand{\generatednumber}{50000}
\newcommand{\xexpr}{( \thisrow{GeV} - 1 )*10} 
\newcommand{\yexpr}{\thisrow{Number} / \generatednumber * \crosssection} 
\begin{tikzpicture}
\begin{axis}[
  xmin = 10
, xmax = 300
, ymin = 10
, ymax = 1060
, xlabel = {Missing $p_T$ [GeV]}
, ylabel = {Cross section per bin [fb]}
]

\addlegendimage{empty legend}
\addlegendentry{$\zeta$}

\addplot[const plot, plotone] 
table [
  x expr = \xexpr 
, y expr = \yexpr
] {1e-7-misspt.dat};
\addlegendentry{$1\times10^{-7}$}

\addplot[const plot, plottwo] 
table [
  x expr = \xexpr 
, y expr = \yexpr
] {5e-8-misspt.dat};
\addlegendentry{$5\times10^{-8}$}

\addplot[const plot, plotthree] 
table [
  x expr = \xexpr
, y expr = \yexpr
] {1e-8-misspt.dat};
\addlegendentry{$1\times10^{-8}$}

\addplot[const plot, plotfour] 
table [
  x expr = \xexpr
, y expr = \yexpr
] {5e-9-misspt.dat};
\addlegendentry{$5\times10^{-9}$}

\addplot[const plot, plotfive] 
table [
  x expr = \xexpr
, y expr = \yexpr
] {1e-9-misspt.dat};
\addlegendentry{$1\times 10^{-9}$}

\end{axis}
\end{tikzpicture}
\caption{%
For $\mu = \unit[100]{GeV}$ and $\zeta = 1\times 10^{-7}$, $5\times 10^{-8}$ most often one neutralino remains undetected, corresponding to an increased missing $p_T$ around the neutralino mass $m_{\chi_1^0}=\unit[102]{GeV}$.
For smaller values of $\zeta$ most of the time both neutralinos leave the detector leading to small missing $p_T$.%
}
\end{xsubcaption}
\hfill
\begin{xsubcaption}
\centering
\tikzsetnextfilename{misspt-200}
\newcommand{\crosssection}{532.6} 
\newcommand{\generatednumber}{50000}
\newcommand{\xexpr}{( \thisrow{GeV} - 1 )*10} 
\newcommand{\yexpr}{\thisrow{Number} / \generatednumber * \crosssection} 
\begin{tikzpicture}
\begin{axis}[
  xmin = 10
, xmax = 300
 , ymin = 3
 , ymax = 36
, xlabel = {Missing $p_T$ [GeV]}
, ylabel = {Cross section per bin [fb]}
]

\addlegendimage{empty legend}
\addlegendentry{$\zeta$}

\addplot[const plot, plotone] 
table [
  x expr = \xexpr 
, y expr = \yexpr
] {1e-7-misspt-200.dat};
\addlegendentry{$1\times10^{-7}$}

\addplot[const plot, plottwo] 
table [
  x expr = \xexpr 
, y expr = \yexpr
] {5e-8-misspt-200.dat};
\addlegendentry{$5\times10^{-8}$}

\addplot[const plot, plotthree] 
table [
  x expr = \xexpr
, y expr = \yexpr
] {1e-8-misspt-200.dat};
\addlegendentry{$1\times10^{-8}$}

\addplot[const plot, plotfour] 
table [
  x expr = \xexpr
, y expr = \yexpr
] {5e-9-misspt-200.dat};
\addlegendentry{$5\times10^{-9}$}

\addplot[const plot, plotfive] 
table [
  x expr = \xexpr
, y expr = \yexpr
] {1e-9-misspt-200.dat};
\addlegendentry{$1\times 10^{-9}$}

\end{axis}
\end{tikzpicture}
\caption{%
For $\mu = \unit[200]{GeV}$ and  $\zeta = 5 \times 10^{-8}$, $1\times 10^{-8}$ most often one neutralino leads to missing $p_T$ around the neutralino mass $m_{\chi_1^0}=\unit[205]{GeV}$.
For smaller values of $\zeta$ most of the time both neutralinos leave the detector undetected.%
}
\end{xsubcaption}
\hfill
\caption{%
Missing $p_T$ at the generator level, defined as the sum of the $p_T$ of all neutrinos and of the neutralinos that decay after they have left the detector.
In the case where only one of the two neutralinos escapes the detector, corresponding to a large value in the second columns in Table~\ref{tab:decay percentage data}, we notice an increased missing $p_T$ around the neutralino mass.
In the case where both neutralino decay outside the detector corresponding to a large value in the first columns in Table~\ref{tab:decay percentage data}, the values for small missing $p_T$ are increased.
This reflects that the neutralinos are mostly back-to-back, due to the absence of other high $p_T$ objects in the decay cascade.%
}
\label{fig:missing pt}
\end{figure}

For smaller values of $\zeta$, one of the neutralinos may decay inside or after the muon system leading to signatures with a larger amount of $\slashed E_T$ as shown in Table~\ref{tab:finstatesouthiggsino}.
We show in Figure~\ref{fig:missing pt} how this possibly gives rise to a missing energy signature as $\zeta$ decreases.
However, this situation would still not be covered by the current direct neutralino- and chargino searches by ATLAS and CMS~\cite{:2012gg,:2012ku,:2012ew} since the lepton reconstruction in these searches requires a track in the pixel detector, often with a certain maximum impact parameter to the primary vertex.
For the same reason that we obtain missing energy from one of the neutralinos decaying outside of the detector, it will also be less probable that the other neutralino decays early enough for the leptons to fulfill such requirements on their inner tracks, as will be illustrated below.

For very small R-parity violation both neutralinos may escape the detector without being observed.
This case cannot be distinguished from the one with R-parity conservation and is for stable higgsinos and heavy coloured particles very challenging to detect~\cite{Baer:2011ec}.
This can be understood just by looking at the typical LHS process shown in Figure~\ref{fig:Higgsino decay chain} and trying to imagine possible signatures in the case of the lightest neutralino being stable.

\begin{table}
\begin{xsubcaption}
\centering
\begin{tabularx}{\textwidth}{rcXc}
    \toprule
    $r\leq \unit[5]{mm}$
  & \multicolumn{3}{c}{Number of neutralino decays before the muon system}
 \\ \cmidrule {2-4}
  & \multicolumn{1}{c}{0}
  & \multicolumn{1}{c}{1}
  & 2
 \\ \midrule
    0
  & \textcolor{red}{R-parity conserving LHS-like}
  & \multicolumn{1}{c}{\textcolor{black!30!green!100}{displaced $\mu^+\mu^-$}}
  & \textcolor{black!30!green!100}{displaced $\mu^+\mu^-$}
 \\ 1
  &
  & \textcolor{blue}{may be covered by existing SUSY searches}
  & \textcolor{black!30!green!100}{displaced $\mu^+\mu^-$}
 \\ 2
  &
  &
  & SM-like
 \\ \cmidrule {2-4}
    Signature
  & \multicolumn{1}{c}{no LHC signature}
  & \multicolumn{1}{c}{possibly large $\slashed E_T$}
  & no or small $\slashed E_T$
 \\ \bottomrule
\end{tabularx}
\caption{%
Legend: Classification of event types depending on the number of neutralino decays inside the detector (columns) and inside the innermost \unit[5]{mm} of the tracker (rows).
The fractions of events quoted in green belong to the type covered by our search strategy, the fraction of events in red cannot be distinguished from the R-parity conserving LHS.
The fraction quoted in blue might be covered by existing SUSY searches (see text), and the black events might be misidentified as SM events.%
}
\label{tab:decay percentage legend}
\end{xsubcaption}
\begin{xsubcaption}
\centering
\begin{tabular}{ccccc}
\toprule
 $\zeta$
  & \multicolumn{4}{c}{$\mu$}
\\ \cmidrule{2-5}
  & 100
  & 200
  & 300
  & 400
\\ \midrule
    $1 \times 10^{-7}$
  & \begin{tabular}{@{}l@{ }l@{ }l@{}} \textcolor{red}{\bf 18.2} & \textcolor{black!30!green!100}{\bf 40.5} & \textcolor{black!30!green!100}{\bf 41.1} \\  & \textcolor{blue}{0.059} & \textcolor{black!30!green!100}{0.182} \\  &  & 0.002 \end{tabular}
  & \begin{tabular}{@{}l@{ }l@{ }l@{}} \textcolor{red}{0.046} & \textcolor{black!30!green!100}{2.59} & \textcolor{black!30!green!100}{\bf 95.6} \\ & \textcolor{blue}{0.004} & \textcolor{black!30!green!100}{1.72} \\ & & 0.012 \end{tabular}
  & \begin{tabular}{@{}l@{ }l@{ }l@{}} \textcolor{red}{0} & \textcolor{black!30!green!100}{0.121} & \textcolor{black!30!green!100}{\bf 96.3} \\ & \textcolor{blue}{0.002} & \textcolor{black!30!green!100}{3.52} \\ & & 0.038 \end{tabular}
  & \begin{tabular}{@{}l@{ }l@{ }l@{}} \textcolor{red}{0} & \textcolor{black!30!green!100}{0.004} & \textcolor{black!30!green!100}{\bf 94.4} \\ & \textcolor{blue}{0} & \textcolor{black!30!green!100}{5.51} \\ & & 0.105 \end{tabular}
 \\ \cmidrule(lr){2-2} \cmidrule(lr){3-3} \cmidrule(lr){4-4} \cmidrule(lr){5-5}
    $5 \times 10^{-8}$
  & \begin{tabular}{@{}l@{ }l@{ }l@{}} \textcolor{red}{\bf 56.1} & \textcolor{black!30!green!100}{\bf 35.5} & \textcolor{black!30!green!100}{8.36} \\ & \textcolor{blue}{0.042} & \textcolor{black!30!green!100}{0.030} \\ & & 0 \end{tabular}
  & \begin{tabular}{@{}l@{ }l@{ }l@{}} \textcolor{red}{4.48} & \textcolor{black!30!green!100}{\bf 28.6} & \textcolor{black!30!green!100}{\bf 66.4} \\ & \textcolor{blue}{0.083} & \textcolor{black!30!green!100}{0.404} \\ & & 0 \end{tabular}
  & \begin{tabular}{@{}l@{ }l@{ }l@{}} \textcolor{red}{0.405} & \textcolor{black!30!green!100}{9.64} & \textcolor{black!30!green!100}{\bf 89.0} \\ & \textcolor{blue}{0.040} & \textcolor{black!30!green!100}{0.928} \\ & & 0.002 \end{tabular}
  & \begin{tabular}{@{}l@{ }l@{ }l@{}} \textcolor{red}{0.028} & \textcolor{black!30!green!100}{2.64} & \textcolor{black!30!green!100}{\bf 95.8} \\ & \textcolor{blue}{0.016} & \textcolor{black!30!green!100}{1.52} \\ & & 0.006 \end{tabular}
 \\ \cmidrule(lr){2-2} \cmidrule(lr){3-3} \cmidrule(lr){4-4}  \cmidrule(lr){5-5}
    $1 \times 10^{-8}$
  & \begin{tabular}{@{}l@{ }l@{ }l@{}} \textcolor{red}{\bf 97.4} & \textcolor{black!30!green!100}{2.58} & \textcolor{black!30!green!100}{0.222} \\ & \textcolor{blue}{0} & \textcolor{black!30!green!100}{0} \\ & & 0 \end{tabular}
  & \begin{tabular}{@{}l@{ }l@{ }l@{}} \textcolor{red}{\bf 82.1} & \textcolor{black!30!green!100}{\bf 16.7} & \textcolor{black!30!green!100}{1.16} \\ & \textcolor{blue}{0.014} & \textcolor{black!30!green!100}{0} \\ & & 0 \end{tabular}
  & \begin{tabular}{@{}l@{ }l@{ }l@{}} \textcolor{red}{\bf 68.7} & \textcolor{black!30!green!100}{\bf 27.8} & \textcolor{black!30!green!100}{3.41} \\ & \textcolor{blue}{0.030} & \textcolor{black!30!green!100}{0.004} \\ & & 0 \end{tabular}
  & \begin{tabular}{@{}l@{ }l@{ }l@{}} \textcolor{red}{\bf 55.3} & \textcolor{black!30!green!100}{\bf 37.0} & \textcolor{black!30!green!100}{7.68} \\ & \textcolor{blue}{0.048} & \textcolor{black!30!green!100}{0.024} \\ & & 0 \end{tabular}
 \\ \cmidrule(lr){2-2} \cmidrule(lr){3-3} \cmidrule(lr){4-4} \cmidrule(lr){5-5}
    $5 \times 10^{-9}$
  & \begin{tabular}{@{}l@{ }l@{ }l@{}} \textcolor{red}{\bf 99.4} & \textcolor{black!30!green!100}{0.643} & \textcolor{black!30!green!100}{0.002} \\ & \textcolor{blue}{0} & \textcolor{black!30!green!100}{0} \\ & & 0 \end{tabular}
  & \begin{tabular}{@{}l@{ }l@{ }l@{}} \textcolor{red}{\bf 95.0} & \textcolor{black!30!green!100}{4.94} & \textcolor{black!30!green!100}{0.083} \\ & \textcolor{blue}{0.008} & \textcolor{black!30!green!100}{0} \\ & & 0 \end{tabular}
  & \begin{tabular}{@{}l@{ }l@{ }l@{}} \textcolor{red}{\bf 90.4} & \textcolor{black!30!green!100}{9.23} & \textcolor{black!30!green!100}{0.323} \\ & \textcolor{blue}{0.008} & \textcolor{black!30!green!100}{0} \\ & & 0 \end{tabular}
  & \begin{tabular}{@{}l@{ }l@{ }l@{}} \textcolor{red}{\bf 85.4} & \textcolor{black!30!green!100}{\bf 13.9} & \textcolor{black!30!green!100}{0.709} \\ & \textcolor{blue}{0.012} & \textcolor{black!30!green!100}{0} \\ & & 0 \end{tabular}
 \\ \cmidrule(lr){2-2} \cmidrule(lr){3-3} \cmidrule(lr){4-4} \cmidrule(lr){5-5}
    $1 \times 10^{-9}$
  & \begin{tabular}{@{}l@{ }l@{ }l@{}} \textcolor{red}{\bf 99.9} & \textcolor{black!30!green!100}{0.018} & \textcolor{black!30!green!100}{0} \\ & \textcolor{blue}{0} & \textcolor{black!30!green!100}{0} \\ & & 0 \end{tabular}
  & \begin{tabular}{@{}l@{ }l@{ }l@{}} \textcolor{red}{\bf 99.8} &  \textcolor{black!30!green!100}{0.204} &  \textcolor{black!30!green!100}{0} \\ & \textcolor{blue}{0} &  \textcolor{black!30!green!100}{0} \\ & & 0 \end{tabular}
  & \begin{tabular}{@{}l@{ }l@{ }l@{}} \textcolor{red}{\bf 99.6} & \textcolor{black!30!green!100}{0.391} & \textcolor{black!30!green!100}{0} \\ & \textcolor{blue}{0.002} & \textcolor{black!30!green!100}{0} \\ & & 0 \end{tabular}
  & \begin{tabular}{@{}l@{ }l@{ }l@{}} \textcolor{red}{\bf 99.4} & \textcolor{black!30!green!100}{0.633} & \textcolor{black!30!green!100}{0.002} \\ & \textcolor{blue}{0.002} & \textcolor{black!30!green!100}{0} \\ & & 0 \end{tabular}
 \\ \bottomrule
\end{tabular}
\caption{%
Generator level fractions in \% of neutralinos decaying inside and outside of the detector as well as in its innermost part.
The meaning of the positions in the subtables and the color code is explained in Legend~(\subref{tab:decay percentage legend}).
Numbers larger than \unit[10]{\%} are bold.
A zero entry means that $\leq \unit[0.001]{\%}$ of the decays happen in this channel.
The tendency of lighter higgsinos with smaller R-parity violation to decay outside the detector follows from relation~\eqref{zeta from neutralino properties}.%
}
\label{tab:decay percentage data}
\end{xsubcaption}
\caption{%
Fractions of neutralino decays occuring either within a radius of \unit[5]{mm}, inside the muon system or outside the detector, depending on the $\zeta$ and $\mu$ parameter.
In Legend~(\subref{tab:decay percentage legend}) we explain the color code and the meaning of the positions in Table~(\subref{tab:decay percentage data}).%
}
\label{tab:decay percentage}
\end{table}

We illustrate in Table~\ref{tab:decay percentage} how often the different situations of none, one or two of the neutralinos, produced in processes like in Figure~\ref{fig:Higgsino decay chain}, decaying inside the detector occurs depending on the RPV and the higgsino mass.
We also check how often the events would give inner tracks, possibly with a small impact parameter to the primary vertex, since this is a common requirement on lepton objects in existing searches.
Table~\ref{tab:decay percentage} illustrates:
\begin{itemize}
\item For decreasing $\zeta$, we approach a situation indistinguishable from the case of R-parity conserving LHS (see numbers in red).
\item The second column in each subtable shows that when we can have large missing transverse energy in the event due to one of the neutralinos decaying outside of the detector, very few events will have the other neutralino decaying sufficiently close to the primary vertex for the event to be seen in existing searches for direct production of charginos and neutralinos (see numbers in blue).
\item We also see that for a given $\zeta$, increasing higgsino mass will lead to more decays inside the detector.
\end{itemize}

The leptonic decays of at least one of the neutralinos inside the detector lead to an opposite-sign dimuon signature, which will be exploited in the present work.
We will focus on events with two opposite sign muons originating either in a secondary vertex in the tracker, far away from the primary interaction point, or having no associated track at all, being identified solely by the muon system.
The muon identification and reconstruction process applied is described in~\cite{Bobrovskyi:2011vx} and will be briefly summarized in Section~\ref{Sec:Analysis}.

\subsection{Benchmark points}

\begin{table}
\centering
\begin{tabular}{rrrrr}
    \toprule
    \multicolumn{1}{c}{higgsino}
  & \multicolumn{4}{c}{$\mu$}
 \\ \cmidrule{2-5}
  & 100
  & 200
  & 300
  & 400
 \\ \midrule
    $\chi_2^0$
  & 106
  & 209
  & 311
  & 413
 \\ $\chi_1^\pm$
  & 104
  & 207
  & 309
  & 411
 \\ $\chi_1^0$
  & 102
  & 205
  & 307
  & 408
 \\ \bottomrule
\end{tabular}
\caption{Mass spectrum of light higgsinos in our four benchmark models with a higgsino mass parameter $\mu$ between \unit[100]{GeV} and \unit[400]{GeV}.
All masses are given in units of GeV.}
\label{tab:higgsino masses}
\end{table}
\begin{table}
\hfill
\begin{xsubcaption}
\begin{tabular}{rd{4.1}d{2.1}d{3.2}d{3.2}}
    \toprule
    \multicolumn{1}{c}{ }
  & \multicolumn{4}{c}{$\mu$}
 \\ \cmidrule{2-5}
  & \multicolumn{1}{c}{100}
  & \multicolumn{1}{c}{200}
  & \multicolumn{1}{c}{300}
  & \multicolumn{1}{c}{400}
 \\ \midrule
    $\chi^0_1 \chi^+_1$
  & 1640
  & 121
  & 22.8
  & 6.28
 \\ $\chi^0_2 \chi^+_1$
  & 1530
  & 116
  & 22.2
  & 6.15
 \\ $\chi^-_1 \chi^+_1$
  & 1300
  & 94.8
  & 17.2
  & 4.58
 \\ $\chi^0_1 \chi^-_1$
  & 918
  & 55.9
  & 9.23
  & 2.29
 \\ $\chi^0_2 \chi^-_1$
  & 851
  & 53.6
  & 8.94
  & 2.24
 \\ $\chi^0_1 \chi^0_2$
  & 1410
  & 91.3
  & 16.1
  & 4.19
 \\ \midrule
    $\sigma^\text{tot}$
  & 7649
  & 532.6
  & 96.47
  & 25.73
 \\ $\mathcal L^\text{gen}_\text{min}$
  & 18.3
  & 93.9
  & 518
  & 1940
 \\ $\mathcal L^\text{gen}_\text{max}$
  & 565
  & 263
  & 1450
  & 5440
 \\ \bottomrule
\end{tabular}
\caption{$\mathcal L = \unit[8]{TeV}$.}
\label{tab:signal cross sections 8 TeV}
\end{xsubcaption}
\hfill
\begin{xsubcaption}
\begin{tabular}{rrrd{3.1}d{2}}
    \toprule
    \multicolumn{1}{c}{ }
  & \multicolumn{4}{c}{$\mu$}
 \\ \cmidrule{2-5}
  & \multicolumn{1}{c}{100}
  & \multicolumn{1}{c}{200}
  & \multicolumn{1}{c}{300}
  & \multicolumn{1}{c}{400}
 \\ \midrule
    $\chi^0_1 \chi^+_1$
  & 3350
  & 293
  & 66.0
  & 21.9
 \\ $\chi^0_2 \chi^+_1$
  & 3130
  & 282
  & 64.3
  & 21.5
 \\ $\chi^-_1 \chi^+_1$
  & 2770
  & 246
  & 53.9
  & 17.4
 \\ $\chi^0_1 \chi^-_1$
  & 2090
  & 158
  & 32.0
  & 9.72
 \\ $\chi^0_2 \chi^-_1$
  & 1950
  & 152
  & 31.2
  & 9.54
 \\ $\chi^0_1 \chi^0_2$
  & 3030
  & 240
  & 51.0
  & 16.2
 \\ \midrule
    $\sigma^\text{tot}$
  & 16320
  & 1371
  & 298.4
  & 96.26
 \\ $\nicefrac{\sigma^\text{tot}_{14}}{\sigma^\text{tot}_8}$
  & 2.1
  & 2.6
  & 3.1
  & 3.7
 \\ \bottomrule \\
\end{tabular}
\caption{$\mathcal L = \unit[14]{TeV}$.}
\label{tab:signal cross sections 14 TeV}
\end{xsubcaption}
\hfill\strut
\caption{Partial and total NLO production cross sections for our benchmark models at \unit[8]{TeV} and \unit[14]{TeV} LHC in units of fb.
The minimal and maximal (depending on $\zeta$) integrated luminosity corresponding to the generated number of events at \unit[8]{TeV} for each model is given in units of fb$^{-1}$.}
\label{tab:prodxsec}
\end{table}

In models with hybrid gauge-gravity mediated supersymmetry breaking it is possible to realize a  spectrum with higgsino masses around the electroweak scale, Higgs partners around the TeV scale and all other particles at the multi TeV scale~\cite{Brummer:2011yd}.

The gravity mediated higgsino parameter $\mu$ of the superpotential and hence the lightest neutralinos and charginos can be of order \unit[100]{GeV}.
The existing lower bound on the chargino mass of roughly \unit[95]{GeV} for degenerate spectra comes from LEP~\cite{lepsusywg:01-03.1, Beringer:1900zz}.
In this study we have chosen four benchmark points for which we have varied the higgsino mass parameter $\mu$ in three steps from the lower bound of about \unit[100]{GeV} to \unit[400]{GeV}, see Table~\ref{tab:higgsino masses}.
The masses of the MSSM Higgs particles are in this case set by the CP-odd Higgs mass parameter $m_A$, which we have taken to be \unit[800]{GeV}.
All other particles are governed by the gauge mediated parameters $m_0$ and $m_\onehalf$ which are chosen to be \unit[3]{TeV}, putting them out of reach of the LHC.

In all our benchmark points the Higgs mass is around \unit[125]{GeV}, in agreement with the observed Higgs-like resonance at the LHC~\cite{:2012gk, :2012gu}.
Furthermore, while the LHCb result of an excess in the search for the rare decay $B_s^0\rightarrow\mu^+\mu^-$ \cite{:2012ct} excludes many models with large $\tan\beta$~\cite{Altmannshofer:2012ks}, the LHS is unaffected by this constraint due to the large mass splitting between the $\mu$ parameter and the squark masses.

We have used these parameter choices as input values for a full RGE calculation performed with \software{softsusy}.
As expected the production cross sections for all supersymmetric particles except the light higgsino states are negligible.
The higgsino production cross sections for the four benchmark points are listed in Table~\ref{tab:prodxsec}.

\subsection{Background}

\begin{table}
\centering
\begin{tabular}{rccccc}
    \toprule
  & \multicolumn{1}{c}{$t\bar t$}
  & \multicolumn{1}{c}{$Z\rightarrow\mu\mu$}
  & \multicolumn{1}{c}{$WW$}
  & \multicolumn{1}{c}{$WZ$}
  & \multicolumn{1}{c}{$ZZ$}
  \\
  & \multicolumn{1}{c}{\cite{Melnikov:2009dn}}
  & \multicolumn{1}{c}{\cite{Campbell:2003hd}}
  & \multicolumn{1}{c}{\cite{Campbell:2011bn}}
  & \multicolumn{1}{c}{\cite{Campbell:2011bn}}
  & \multicolumn{1}{c}{\cite{Campbell:2011bn}}
 \\ \midrule
    $\sigma_\text{NLO}$
  & 183
  & 536
  & 57.25
  & 18.55
  & 7.92
 \\ $\mathcal L^\text{gen}$
  & 196
  & 167
  & 360
  & 306
  & 143
 \\ \bottomrule
\end{tabular}
\caption{NLO production cross sections for the relevant background processes in units of pb at an energy of \unit[8]{TeV} as well as the integrated luminosity corresponding to the generated number of events in units of fb$^{-1}$.}
\label{tab:background production cross sections}
\end{table}

The SM processes that dominate the dimuon channel are:
\begin{itemize}
\item $\gamma^* / Z^* \rightarrow \mu^+ \mu^-$
\item $t\bar t$
\item $V^* V^*$, where $V = W, Z$.
\end{itemize}
In Table~\ref{tab:background production cross sections} we give the next-to-leading order (NLO) cross section for the processes that we have simulated for our study. As we will see, these will be efficiently removed by the requirement of a secondary vertex.

In our analysis, we will require the muons to be isolated, which efficiently removes leptons originating in jets, and we further remove possible contributions from displaced $b$ quarks by a sufficiently large cut on tracks in the inner radius.
For low background levels, however, other background sources might come into play. These are:
\begin{itemize}
\item cosmic muons,
\item pion and kaon decays in flight,
\item hadronic punch-throughs,
\item pileup.
\end{itemize}
An estimation of such contributions to our background has to be done with real LHC data, and is beyond the scope of this work.
We argue here that most of this background, should it contribute, can be removed without significant loss of signal.
Cosmic muons can be vetoed against by using the timing information, as discussed in~\cite{Bobrovskyi:2011vx}, or a cut on back-to-back muons.
Punch-throughs are also not simulated in Delphes but should in principle be possible to veto since in this case the muon would be associated with a jet.
Most of any possible contribution to displaced muons from decay in flight should be removed by our high $p_T$ requirement on muons.
Pileup was estimated in a partly similar analysis to give a systematic uncertainty in the event selection efficiency of \unit[2]{\%}~\cite{:2012ae}.
The displacement due to pileup is in general much smaller than the secondary vertices we are expecting.
Therefore, such a background can be reduced by increasing the minimal impact parameter value required, which in our case of larger displacements would not lead to a large decrease in signal efficiency.

In the following we therefore neglect these backgrounds to our displaced muon channel.
However, as will be described in Section~\ref{sec:result}, we will in our statistical analysis allow some margin for systematic uncertainty in case of vanishing estimated background levels by requiring our predicted signal to amount to a certain number of observed events.


\subsection{Analysis}
\label{Sec:Analysis}

\subsubsection{Tools and settings}

All Monte Carlo samples were generated using \software{madgraph} 4.4.44~\cite{Alwall:2007st} interfaced with \software{Pythia} 6.4.22~\cite{Sjostrand:2006za}.
We have chosen parton distribution functions given by CTEQ6L1~\cite{Pumplin:2002vw} and have used a modified version of \software{softsusy} 3.2.4~\cite{Allanach:2001kg} for the calculation of the R-parity violating decays according to the formulas~\eqref{Neutralino_Decay_Widths}.
All other signal decays were calculated using \software{sdecay}~\cite{Muhlleitner:2004mka}.

The generic detector simulation \software{delphes} 1.9~\cite{Ovyn:2009tx}, tuned to the CMS detector%
\footnote{As discussed in Section~\ref{sec: search strategy}, our results would be similar for the ATLAS detector.},
was used in order to account for effects of event reconstruction at the detector level.
The finite radial size of the detector important for studies with secondary vertices has been taken into account for the case of muons following the muon reconstruction procedure described in~\cite{Bobrovskyi:2011vx}.
We have reconstructed displaced muons with and without inner tracks by using the information at the event generator level to deduce where they come into existence. The generator level particles and the detector level objects are matched including the distance in pseudorapidity and azimuthal angle.
In the reconstruction process we have assumed that a muon can be reconstructed as long as it is created before the muon system and that an inner track can be reconstructed as long as the muon is created within the first third of the tracker chamber.

After the muon reconstruction process we are left with two non-overlapping classes of muon-objects: First what we will refer to as \emph{chamber muons}, that are identified solely by the muon system, and second the \emph{tracker muons} having in addition the information about their origin from the associated inner track%
\footnote{In experimental jargon, these can also be referred to at \emph{standalone} and \emph{combined} muons, respectively.}.
The simulated muon reconstruction efficiency is close to \unit[80]{\%} in the inner parts of the detector and drops rapidly if the muon is created in the muon system.

\subsubsection{Cuts}

\begin{table}
\centering
\begin{tabular}{r@{ }ld{4.1}d{3.0}d{3}d{3}d{3}}
    \toprule
    \multicolumn{2}{c}{cuts}
  & \multicolumn{1}{c}{$t\bar t$} 
  & \multicolumn{1}{c}{$Z\rightarrow\mu\mu$} 
  & \multicolumn{1}{c}{$WW$} 
  & \multicolumn{1}{c}{$WZ$} 
  & \multicolumn{1}{c}{$ZZ$} 
 \\ \midrule
    $N( \mu )$&$\ge2$
  & 3057 
  & 397 
  & 410 
  & 361 
  & 283 
 \\ \cmidrule{3-7}
    \multicolumn{2}{c}{Class~1}
  & 2177 
  & 385 
  & 352 
  & 297 
  & 256 
 \\ $m_\text{inv}(\mu^+\mu^-)$&$>\unit[5]{GeV}$
  & 1761 
  & 385 
  & 351 
  & 297 
  & 256 
 \\ $d(\text{Vertex})$&$>\unit[5]{mm}$
  & 11.2 
  & 0
  & 0.369 
  & 0.281 
  & 0.189 
 \\ $\Delta d(\text{Vertex})_{ij}$&$<\unit[1]{mm}$
  & 0
  & 0
  & 0
  & 0
  & 0
 \\ \cmidrule{3-7}
    \multicolumn{2}{c}{Class~2}
  & 0
  & 0
  & 0
  & 0
  & 0
 \\ $m_\text{inv}(\mu^+\mu^-)$&$>\unit[5]{GeV}$
  & 0
  & 0
  & 0
  & 0
  & 0
 \\ \midrule
    \multicolumn{2}{c}{Total}
  & 0
  & 0
  & 0
  & 0
  & 0
 \\ \bottomrule
\end{tabular}
\caption{
Cutflow for the main SM dimuon background in units of \unit{fb}.
}
\label{tab:Background cutflow}
\end{table}

\begin{table}
\centering
\begin{tabular}{r@{ }ld{1.3}d{1.3}d{1.3}d{1.3}}
    \toprule
    \multicolumn{2}{c}{cuts}
  & \multicolumn{2}{c}{100}
  & \multicolumn{2}{c}{200}
 \\ \cmidrule(r){3-4} \cmidrule(l){5-6}
  &
  & \multicolumn{1}{c}{$1 \times 10^{-8}$} 
  & \multicolumn{1}{c}{$5 \times 10^{-9}$} 
  & \multicolumn{1}{c}{$1 \times 10^{-8}$} 
  & \multicolumn{1}{c}{$5 \times 10^{-9}$} 
 \\ \midrule
    $N( \mu )$&$\ge 2$
  & 4.26 
  & 1.64 
  & 3.50 
  & 0.82 
 \\ \cmidrule(r){3-4} \cmidrule(l){5-6}
    \multicolumn{2}{c}{Class~1}
  & 0.219 
  & 0.109 
  & 0.394 
  & 0.096 
 \\ $m_\text{inv}(\mu^+\mu^-)$&$>\unit[5]{GeV}$
  & 0.219 
  & 0.109 
  & 0.394 
  & 0.096 
 \\ $d( \text{Vertex} )$&$> \unit[5]{mm}$
  & 0.219 
  & 0.109 
  & 0.394 
  & 0.096 
 \\ $\Delta d( \text{Vertex} )_{ ij }$&$< \unit[1]{mm}$
  & 0.219 
  & 0.109 
  & 0.341 
  & 0.075 
 \\ \cmidrule(r){3-4} \cmidrule(l){5-6}
    \multicolumn{2}{c}{Class~2}
  & 3.77 
  & 1.31 
  & 2.49 
  & 0.692 
 \\ $m_\text{inv}(\mu^+\mu^-)$&$>\unit[5]{GeV}$
  & 3.66 
  & 1.31 
  & 2.49 
  & 0.692 
 \\ \midrule
  \multicolumn{2}{c}{Total}
  & 3.88 
  & 1.42 
  & 2.83 
  & 0.767 
 \\ \bottomrule
\end{tabular}
\caption{Cutflow in units of fb for the two lighter benchmark points ($\mu=100$, 200 GeV) with the two most relevant of the analysed values of R-parity breaking $\zeta$.
}
\label{tab:Signal cutflow}
\end{table}

We focus our search solely on the muon objects as we assume that we can trust the detector simulation results in this case even in the presence of secondary vertices.
We aim to reconstruct clear signatures where one of the neutralinos decays into a muon and a $W$~boson, which in turn decays into another muon and a neutrino.
Thus, we demand two opposite sign muons that are either chamber muons or tracker muons with a secondary vertex far away from the primary interaction point.
We remind the reader that displaced muons are not expected in SM processes giving two isolated muons, and we will show in this section that the requirement of large displacements efficiently removes the SM dimuon background.
Furthermore, we require that the invariant mass of the two-muon system is not too small, thus suppressing back-to-back signals.
This reduces not only the SM background, but also helps to decrease the background of cosmic muons.

Only \software{delphes} muon objects with a $p_T$ larger than \unit[10]{GeV} are passed in our additional muon reconstruction processes described above.
We demand that the secondary vertex of the muons lies before the muon system, meaning $r(\text{Vertex}) < \unit[4]{m}$ and $z(\text{Vertex}) < \unit[6]{m}$, where $r$ and $z$ are the radial coordinate perpendicular to the beam and the coordinate parallel to the beam, respectively.
We assume that tracks can be reconstructed reliably as long as the secondary vertex lies inside the cylinder defined by $r(\text{Vertex}) < \unit[40]{cm}$ and $z(\text{Vertex}) < \unit[1.3]{m}$.
The cut on the pseudorapidity for muons is taken to be $\eta < 2.5$.
In addition to the isolation requirements in \software{delphes}, we only select muons that have no overlap with jet objects in angular coordinates $\Delta R > 0.1$, where
\begin{align}
    \Delta R
  = \sqrt{ \Delta \phi^2 + \Delta \eta^2 }
\ ,
\end{align}
and $\Delta \phi$ and $\Delta \eta$ are the appropriate angle differences between the muon and the jet in the usual detector coordinates.
Then we perform a selection cut on the total number of muons (chamber and tracker muons) in the event:
\begin{itemize}
 \item $N( \text{muons} ) \ge 2$.
\end{itemize}
We define two event classes:
\begin{itemize}
 \item Class~1: the event contains exactly two tracker muons, \emph{i.e.}\ $N( \text{tracker muons} ) = 2$, with opposite charge.
 \item Class~2: the event does not fulfill the conditions for Class~1 and contains exactly two chamber muons, \emph{i.e.}\ $N(\text{chamber muons}) = 2$, of opposite charge.
\end{itemize}
As the amount of R-parity breaking decreases, more and more events will fall into the second class.
In accordance with the description of the signal presented above, we implement the following cuts on the Class~1 events:
\begin{itemize}
 \item $m_\text{inv}( \mu^+ \mu^- ) > \unit[5]{GeV}$:
  We compute the invariant mass of the muon pair and demand that it is larger than \unit[5]{GeV}.
 \item $d(\text{Vertex}) > \unit[5]{mm}$:
  Each of the tracks associated with the two tracker muons should have a vertex which is further than \unit[5]{mm} away from the primary vertex. This value is approximately one order of magnitude larger than the resolution of the inner tracker~\cite{Bayatian:2006zz,Aad:2009wy}.
 \item $\Delta d(\text{Vertex})_{ij} < \unit[1]{mm}$:
  The distance between the two track vertices should be less than \unit[1]{mm}, in order to capture events where both muons originate in the same secondary vertex.
\end{itemize}
If the event fails one of the above cuts and fulfills the criteria for Class~2 events it is classified as Class~2:
\begin{itemize}
  \item $m_\text{inv}( \mu^+ \mu^- ) > \unit[5]{GeV}$:
   Also in this case we demand that the invariant mass of the muon pair is larger than \unit[5]{GeV}.
\end{itemize}
As expected, all of the SM background events are removed by the cut on the minimal distance of the vertex from  the primary interaction point and the requirement that the reconstructed secondary vertices are close to each other (see Table~\ref{tab:Background cutflow}).
The LHS model events, however, survive these cuts to such an extent that a signal is detectable (see Table~\ref{tab:Signal cutflow}).

\subsubsection{Mass determination}
\label{sec:mass determination}

\begin{figure}
\small
\renewcommand{\Width}{.385\textwidth}
\renewcommand{\Height}{\Width}
\begin{xsubcaption}[.499\textwidth]
\tikzsetnextfilename{mass-edge-large}
\newcommand{\normx}[1]{ ( x / sqrt( 2 ) / #1 )} 
\newcommand{\normedge}[2]{ ( ( x - #1 ) / sqrt( 2 ) / #2 )} 
\newcommand{\edge}[2]{ sqrt( #2 ) / 2 * ( x * erf( \normx{#2} ) - x * erf( \normedge{#1}{#2} ) + sqrt( 2 / 3.1 ) * #2 * ( exp( - \normx{#2}^2 ) - exp( - \normedge{#1}{#2}^2 ) ) ) } 
\begin{tikzpicture}%
\begin{axis}[
  xmin = 0
, xmax = 80
, ymin = 0
, xlabel = {Invariant mass of $\mu^+\mu^-$ [GeV]}
, ylabel = {Number of events per bin}
, legend style = { legend pos = north west }
, every axis legend/.append style={nodes={right}}
]

\addlegendimage{empty legend}
\addlegendentry{\#\quad\quad edge }

\addplot[const plot, forget plot, plotone] table [
  x expr = \thisrow{GeV}*260/100
, y expr = \thisrow{Number}
] {mass-edge-5e-8.dat};
\addplot gnuplot [domain=0:80, samples = 100, plotone] {\edge{65.85}{3}};
\addlegendentry{1682\quad63.7}

\addplot[const plot, forget plot, plottwo] table [
  x expr = \thisrow{GeV}*198/100
, y expr = \thisrow{Number}
] {mass-edge-1e-7.dat};
\addplot gnuplot [domain=0:80, samples = 100, plottwo] {\edge{63.35}{1.4786}};
\addlegendentry{1405\quad63.9}

\addplot[const plot, forget plot, plotthree] table [
  x expr = \thisrow{GeV}*134/100
, y expr = \thisrow{Number}
] {mass-edge-5e-7.dat};
\addplot gnuplot [domain=0:80, samples = 100, plotthree] {\edge{63.7}{0.337}};
\addlegendentry{981\phantom{5}\quad 63.7}

\end{axis}
\begin{axis}[
  xmin=0
, xmax=80
, axis y line = none
, axis x line = top
, xtick = {62.4}
, xticklabel = 62.4
, every outer x axis line/.style={\empty}
]
\addplot[draw= none] {x};
\draw[plotfour] ({axis cs:62.4,0}|-{rel axis cs:0,1}) -- ({axis cs:62.4,0}|-{rel axis cs:0,0});
\end{axis}
\end{tikzpicture}%
\caption{Large total number of events (\#).}
\label{fig:mass-edge large}
\end{xsubcaption}
\hfill
\begin{xsubcaption}[.499\textwidth]
\tikzsetnextfilename{mass-edge-small}
\pgfplotsset{
 every axis/.append style = {
  , legend style = { draw = none, text opacity = 1, fill opacity = .5 }
  }
 }
\newcommand{\normx}[1]{ ( x / sqrt( 2 ) / #1 )} 
\newcommand{\normedge}[2]{ ( ( x - #1 ) / sqrt( 2 ) / #2 )} 
\newcommand{\edge}[2]{ sqrt( #2 ) / 2 * ( x * erf( \normx{#2} ) - x * erf( \normedge{#1}{#2} ) + sqrt( 2 / 3.1 ) * #2 * ( exp( - \normx{#2}^2 ) - exp( - \normedge{#1}{#2}^2 ) ) ) } 
\newcommand{\xmax}{80}
\newcommand{\massedge}{62.4}
\begin{tikzpicture}%
\begin{axis}[
  xmin = 0
, xmax = \xmax
, ymin = 0
, xlabel = {Invariant mass of $\mu^+\mu^-$ [GeV]}
, ylabel = {Number of events per bin}
, legend style = { legend pos = north west }
, every axis legend/.append style={ nodes = right }
]

\tikzset{ 
  opaquestyle/.style =  { fill opacity = .5, draw = none, fill}
}

\pgfplotsset{
  legend image code/.code={%
    \draw[#1, opaquestyle] (0cm,-0.1cm) rectangle (0.6cm,0.1cm);
    \draw[#1] (0cm,0cm) -- (0.6cm,0cm);
  }
}

\addlegendimage{empty legend}
\addlegendentry{\#\quad edge }

\addplot[const plot, opaquestyle, plotthree, forget plot] table [
  x expr = \thisrow{GeV}*70/100
, y expr = \thisrow{Number}
] {mass-edge-1e-8.dat};
\addplot gnuplot [domain=0:80, samples = 100, plotthree] {\edge{64.75}{0.0008167}};
\addlegendentry{73\quad64.7}

\addplot[const plot, opaquestyle, plottwo, forget plot] table [
  x expr = \thisrow{GeV}*70/100
, y expr = \thisrow{Number}
] {mass-edge-1e-8-2.dat};
\addplot gnuplot [domain=0:80, samples = 100, plottwo] {\edge{63.84}{0.000722}};
\addlegendentry{50\quad63.8}

\addplot[const plot, opaquestyle, plotone, forget plot] table [
  x expr = \thisrow{GeV}*72/100
, y expr = \thisrow{Number}
] {mass-edge-5e-9-2.dat};
\addplot gnuplot [domain=0:80, samples = 100, plotone] {\edge{67.3}{0.000443}};
\addlegendentry{26\quad67.3}

\end{axis}
\begin{axis}[
  xmin=0
, xmax= \xmax
, axis y line = none
, axis x line = top
, xtick = {\massedge}
, xticklabel = \massedge
, every outer x axis line/.style={\empty}
]
\addplot[draw= none] {x};
\draw[plotfour] ({axis cs:\massedge,0}|-{rel axis cs:0,1}) -- ({axis cs:\massedge,0}|-{rel axis cs:0,0});
\end{axis}
\end{tikzpicture}%
\caption{Small total number of events (\#).}
\label{fig:mass-edge small}
\end{xsubcaption}
\caption{Examples of the mass-edge reconstruction when the higgsino mass is $m_{\chi_1^0} = \unit[102]{GeV}$ (benchmarks $\mu = \unit[100]{GeV}$), so that the theoretical value for the edge in the dimuon invariant mass is \unit[62.4]{GeV}.
The values in the legend show that already a small total number of events (\#) are sufficient in order to reconstruct the mass edge with an error around \unit[2]{GeV}.
}
\label{fig:mass-edge}
\end{figure}

Rejecting the background-only hypothesis is only one contribution to the degree of belief that new physics has been discovered.
Whether the signal hypothesis is a plausible one should be tested in other ways as well.
Here, we show that the chosen signature allows for the determination of the neutralino mass via the well-known mass edge method~\cite{Hinchliffe:1996iu,Allanach:2000kt}.
The mass edge in the dimuon invariant mass distribution is to leading order determined by
\begin{align}
    m_{ l l }^2
  = m_{ \chi^0_1 }^2 - m_W^2
  + \mathcal O \left( \frac{ m_l }{ m_{ \chi_1^0 } } \right)^2
\ .
\end{align}
Following~\cite{CMS-PAS-SUS-09-002} we fold the phase space function with a Gaussian to model the mass edge:
\begin{align}
    T( m_{ l l } )
  = \frac{ 1 }{ \sqrt{ 2 \pi \sigma } } \int_0^{ m_\text{cut}} dy\ y
    \exp \left( - \frac{ 1 }{ 2 } \left( \frac{ m_{ l l } - y }{ \sigma } \right)^2 \right)
\ ,
\end{align}
where the endpoint $m_\text{cut}$ and the height of the triangle $\sigma$ are the free parameters to be fitted to the dilepton invariant mass $m_{ l l }$ distribution to reconstruct the dimuon mass edge.
We implemented this mass edge formula in the \software{minuit} class of the \software{root} package.

For this method to work, a sufficiently large sample of signal events is needed.
In Figure~\ref{fig:mass-edge} we show examples of the mass edge reconstruction for different numbers of observed events in the case of our benchmarks model with $\mu = \unit[100]{GeV}$.
We conclude from Figure~\ref{fig:mass-edge small} that a total number of events between 26 and 50 should give an accurate estimate of the higgsino mass.

\subsection{Result}\label{sec:result}

\begin{table}
\centering
\begin{tabular}{ r d{2.3} @{ $\pm$ } d{1.3} d{2.3} @{ $\pm$ } d{1.3} d{1.3} @{ $\pm$ } d{1.3} d{1.4} @{ $\pm$ } d{1.4} }
\toprule
    \multicolumn{1}{c}{$\zeta$}
  & \multicolumn{8}{c}{$\mu$}
 \\ \cmidrule{2-9}
  & \multicolumn{2}{c}{100}
  & \multicolumn{2}{c}{200}
  & \multicolumn{2}{c}{300}
  & \multicolumn{2}{c}{400}
 \\ \midrule
 $5\times10^{-8}$
  & 90 & 2 
  & 25.6 & 0.5 
  & 4.9 & 0.1 
  & 1.17 & 0.03 
 \\ $1\times10^{-8}$
  & 3.9 & 0.5 
  & 2.8 & 0.2 
  & 1.07 & 0.05 
  & 0.39 & 0.01 
 \\ $5\times10^{-9}$
  & 1.4 & 0.3 
  & 0.77 & 0.09 
  & 0.27 & 0.02 
  & 0.105 & 0.007 
 \\ $1\times10^{-9}$
  & 0.028 & 0.007 
  & 0.023 & 0.009 
  & 0.014 & 0.003 
  & 0.0037 & 0.0008 
 \\ \bottomrule
\end{tabular}
\caption{%
Signal cross sections after cuts for all benchmark models and different values of the RPV parameter $\zeta$, in units of \unit{fb}.
The errors are Poisson errors and the center-of-mass energy \unit[8]{TeV}.%
\label{tab:signal cross sections after cuts}
}
\end{table}

The signal cross sections after cuts for all our LHS benchmark models are given in Table~\ref{tab:signal cross sections after cuts}. The SM background is removed by our cuts, as shown in Table~\ref{tab:Background cutflow}.

When dealing with very low background levels, a Gaussian approximation may not be adequate, and one should assume the number of events to be Poisson distributed. Under the null hypothesis of $B$ background events, the probability of observing $N$ or fewer events is then
\begin{align}
    P( N; B )
  = \sum_{ k = 0 }^N \frac{ B^k }{ k! } e^{ -B }
\ ,
\end{align}
given that the expectation value $B$ is the true mean.
We denote the expected number of events predicted by the model with $S$.
To estimate the integrated luminosity%
\footnote{The integrated luminosity is $\mathcal L = \sigma N$ where $\sigma$ is the cross section and $N$ the number of events.}
needed for a 5~sigma detection, one can require that there is a value for the minimum number of observed events $N_\text{obs} = N + 1$ such that $1 - P(N; B) < \unit[2.9 \times 10^{-5}]{\%}$, corresponding to 5 standard deviations in the case of a one-sided Gaussian.
In addition, $N_\text{obs}$ has to satisfy $( 1 - P(N; S + B) )$ being larger than some probability $P_\text{obs}$ for observation  under the hypothesis of our model.%
\footnote{For a discussion of the statistical measures used for this kind of study, see Appendix B of~\cite{Gustafsson:2012aj}.}

\begin{figure}
\centering
\small
\tikzsetnextfilename{Discovery-Reach}
\newcommand{\xmin}{1}
\newcommand{\xmax}{100}
\newcommand{\ymin}{1e-9}
\newcommand{\ymax}{2e-8}

\newcommand{\sigmaone}{7649}
\newcommand{\sigmatwo}{532.6}
\newcommand{\sigmathree}{96.47}
\newcommand{\sigmafour}{25.73}

\newcommand{\genevents}{50000}
\newcommand{\geneventsbig}{140000}

\newcommand{\thresholdmin}{5.67}
\newcommand{\thresholdopt}{9.27}
\newcommand{\thresholdmax}{13.1}
\newcommand{\fermibound}{1.17e-8}

\newcommand{\lumi}[4]{ ( #1 *#2 ) / ( #3 * #4 ) }  

\pgfplotsset{
  legend image code/.code={%
    \draw[#1, fill, fill opacity = 0.5] (0cm,-0.1cm) rectangle (0.6cm,0.1cm);
    \draw[#1] (0cm,0cm) -- (0.6cm,00cm);
  }
}

\pgfplotsset{
 fillplot/.style = { fill, fill opacity = 0.5, forget plot }
}

\begin{tikzpicture}

\pgfplotstableread{lum_fb_Pobs50_mu100.dat}\minone
\pgfplotstableread{lum_fb_Pobs90_mu100.dat}\medone
\pgfplotstableread{lum_fb_Pobs99_mu100.dat}\maxone

\pgfplotstableread{lum_fb_Pobs50_mu200.dat}\mintwo
\pgfplotstableread{lum_fb_Pobs90_mu200.dat}\medtwo
\pgfplotstableread{lum_fb_Pobs99_mu200.dat}\maxtwo

\pgfplotstableread{lum_fb_Pobs50_mu300.dat}\minthree
\pgfplotstableread{lum_fb_Pobs90_mu300.dat}\medthree
\pgfplotstableread{lum_fb_Pobs99_mu300.dat}\maxthree

\pgfplotstableread{lum_fb_Pobs50_mu400.dat}\minfour
\pgfplotstableread{lum_fb_Pobs90_mu400.dat}\medfour
\pgfplotstableread{lum_fb_Pobs99_mu400.dat}\maxfour

\pgfplotstablesort[sort key={zeta},sort cmp={float >}]\minonesort{\minone}
\pgfplotstablesort[sort key={zeta},sort cmp={float <}]\maxonesort{\maxone}

\pgfplotstablesort[sort key={zeta},sort cmp={float >}]\mintwosort{\mintwo}
\pgfplotstablesort[sort key={zeta},sort cmp={float <}]\maxtwosort{\maxtwo}

\pgfplotstablesort[sort key={zeta},sort cmp={float >}]\minthreesort{\minthree}
\pgfplotstablesort[sort key={zeta},sort cmp={float <}]\maxthreesort{\maxthree}

\pgfplotstablesort[sort key={zeta},sort cmp={float >}]\minfoursort{\minfour}
\pgfplotstablesort[sort key={zeta},sort cmp={float <}]\maxfoursort{\maxfour}

\pgfplotstablevertcat{\fillone}{\minonesort}
\pgfplotstablevertcat{\fillone}{\maxonesort}

\pgfplotstablevertcat{\filltwo}{\mintwosort}
\pgfplotstablevertcat{\filltwo}{\maxtwosort}

\pgfplotstablevertcat{\fillthree}{\minthreesort}
\pgfplotstablevertcat{\fillthree}{\maxthreesort}

\pgfplotstablevertcat{\fillfour}{\minfoursort}
\pgfplotstablevertcat{\fillfour}{\maxfoursort}

\begin{loglogaxis}[
      xmin = \xmin
 , xmax = \xmax
 , xlabel = {Integrated Luminosity $\mathcal L $ [$\text{fb}^{-1}$] }
 , ymin = \ymin
 , ymax = \ymax
 , ylabel = {Amount of R-parity violation $\zeta$}
 , extra y tick style={ log identify minor tick positions = true }
 , legend style = { legend pos = south west }
]

\addlegendimage{empty legend}
\addlegendentry{$\mu$}

\addplot[ plotone, fill, fill opacity=.5, smooth, forget plot]  table [ x index = 0, y index = 1 ] {\fillone};
\addplot[plotone, smooth] table[ x expr = \thisrow{Lum}, y expr = \thisrow{zeta}]  {\medone};
\addlegendentry{100}

\addplot[ plottwo, fill, fill opacity=.5, smooth , forget plot]  table [ x index = 1, y index = 0 ] {\filltwo};
\addplot[plottwo, smooth] table[ x expr = \thisrow{Lum}, y expr = \thisrow{zeta}]  {\medtwo};
\addlegendentry{200}

\addplot[ plotthree, fill, fill opacity=.5, smooth, forget plot ]  table [ x index = 0, y index = 1 ] {\fillthree};
\addplot[plotthree, smooth] table[ x expr = \thisrow{Lum}, y expr = \thisrow{zeta}]  {\medthree};
\addlegendentry{300}

\addplot[ plotfour, fill, fill opacity=.5, smooth, forget plot ]  table [ x index = 0, y index = 1 ] {\fillfour};
\addplot[plotfour, smooth] table[ x expr = \thisrow{Lum}, y expr = \thisrow{zeta}]  {\medfour};
\addlegendentry{400}

\addplot [ pattern = north east lines ]
coordinates{
 ( \xmin, \fermibound )
 ( \xmax, \fermibound )
 ( \xmax, \ymax )
 ( \xmin, \ymax )
} \closedcycle ;

\node [fill = white] at ( axis cs: 10, 1.5e-8) {Fermi-LAT Excluded};

\end{loglogaxis}

\end{tikzpicture}
\caption{%
Discovery reach with \unit[8]{TeV} center-of-mass energy at the LHC for our four benchmark models.
Each colored band represents a value of $\mu$ and the lower, middle and upper line on each band corresponds to $P_\text{obs} = \unit[50]{\%}$, \unit[90]{\%} and \unit[99]{\%}, respectively.%
}
\label{fig:Discovery Reach}
\end{figure}
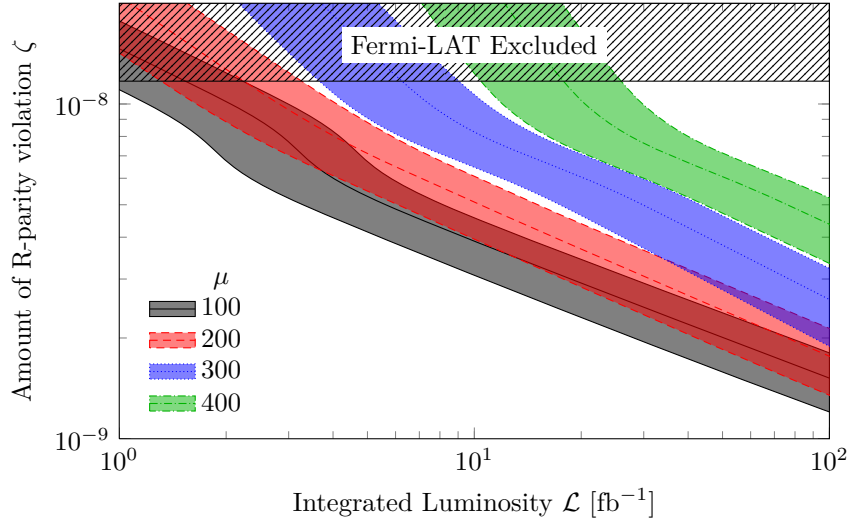

In our case, the expectation is $B = 0$, and in principle any $S>0$ would constitute a signal.
In a real measurement, however, the estimated background will be known only to a limited precision, and we will require $N_\text{obs} \geq 5$ in order to have some margin to allow for systematic uncertainties.
In Figure~\ref{fig:Discovery Reach} we present results assuming a $P_\text{obs}$ of \unit[50]{\%}, \unit[90]{\%} and \unit[99]{\%}.
We see that the integrated luminosities of $\sim \unit[30]{fb^{-1}}$ expected 
in the \unit[8]{TeV} run at the LHC would suffice for discovery of the lightest higgsinos with R-parity violation in the range $\zeta \sim 2 \times 10^{-9} \text{ -- } 2 \times 10^{-8}$, and for $\zeta$ above $6\times10^{-9}$ masses of $\mu = \unit[400]{GeV}$ may be reached.

\begin{figure}
\centering
\small
\tikzsetnextfilename{Mass-Reconstruction}
\newcommand{\xmin}{10}
\newcommand{\xmax}{1000}
\newcommand{\ymin}{1e-9}
\newcommand{\ymax}{2e-8}
\newcommand{\sigmaone}{7649}
\newcommand{\sigmatwo}{532.6}
\newcommand{\sigmathree}{96.47}
\newcommand{\sigmafour}{25.73}
\newcommand{\genevents}{50000}
\newcommand{\geneventsbig}{140000}
\newcommand{\geneventslarge}{4320000}
\newcommand{\thresholdmin}{30}
\newcommand{\thresholdopt}{40}
\newcommand{\thresholdmax}{50}
\newcommand{\fermibound}{1.17e-8}
\newcommand{\lumi}[4]{ ( #1 *#2 ) / ( #3 * #4 ) }  
\pgfplotsset{
 fillplot/.style = { fill, fill opacity = 0.5, forget plot }
}
\pgfplotsset{
  legend image code/.code={%
    \draw[#1, fill, fill opacity = 0.5] (0cm,-0.1cm) rectangle (0.6cm,0.1cm);
    \draw[#1] (0cm,0cm) -- (0.6cm,00cm);
  }
}
\begin{tikzpicture}
\begin{loglogaxis}[ 
  xmin = \xmin
, xmax = \xmax
,  xlabel = { Integrated Luminosity $\mathcal L $ [$\text{fb}^{-1}$]  }
, ymin = \ymin
, ymax = \ymax
, ylabel = { Amount of R-parity violation $\zeta$ }
, extra y tick style={ log identify minor tick positions = true }
, legend style = { legend pos = south west }
]
\addlegendimage{empty legend}
\addlegendentry{$\mu$}

\addplot+[ smooth, plotone, fillplot]
coordinates{
 ( \lumi{\thresholdmin}{\geneventsbig}{\sigmaone}{1647}, 5e-8 )
 ( \lumi{\thresholdmin}{\geneventsbig}{\sigmaone}{71}, 1e-8 ) 
 ( \lumi{\thresholdmin}{\geneventsbig}{\sigmaone}{26}, 5e-9 ) 
 ( \lumi{\thresholdmin}{\geneventslarge}{\sigmaone}{16}, 1e-9 )
 ( \lumi{\thresholdmax}{\geneventslarge}{\sigmaone}{16}, 1e-9 )
 ( \lumi{\thresholdmax}{\geneventsbig}{\sigmaone}{26}, 5e-9 ) 
 ( \lumi{\thresholdmax}{\geneventsbig}{\sigmaone}{71}, 1e-8 )
 ( \lumi{\thresholdmax}{\geneventsbig}{\sigmaone}{1647}, 5e-8 )
} \closedcycle;

\addplot+[ smooth, plotone]
coordinates{
 ( \lumi{\thresholdopt}{\geneventsbig}{\sigmaone}{1647}, 5e-8 )
 ( \lumi{\thresholdopt}{\geneventsbig}{\sigmaone}{71}, 1e-8 )
 ( \lumi{\thresholdopt}{\geneventsbig}{\sigmaone}{26}, 5e-9 )
 ( \lumi{\thresholdopt}{\geneventslarge}{\sigmaone}{16}, 1e-9 )
};
\addlegendentry{100}

\addplot+[smooth, plottwo, fillplot ] coordinates{
 ( \lumi{\thresholdmin}{\genevents}{\sigmatwo}{2406}, 5e-8 )
 ( \lumi{\thresholdmin}{\genevents}{\sigmatwo}{266}, 1e-8 )
 ( \lumi{\thresholdmin}{\genevents}{\sigmatwo}{72}, 5e-9 )
 ( \lumi{\thresholdmin}{\geneventsbig}{\sigmatwo}{6}, 1e-9)
 ( \lumi{\thresholdmax}{\geneventsbig}{\sigmatwo}{6}, 1e-9)
 ( \lumi{\thresholdmax}{\genevents}{\sigmatwo}{72}, 5e-9 )
 ( \lumi{\thresholdmax}{\genevents}{\sigmatwo}{266}, 1e-8 )
 ( \lumi{\thresholdmax}{\genevents}{\sigmatwo}{2406}, 5e-8 )
} \closedcycle;

\addplot+[smooth, plottwo ] coordinates{
 ( \lumi{\thresholdopt}{\genevents}{\sigmatwo}{2406}, 5e-8 )
 ( \lumi{\thresholdopt}{\genevents}{\sigmatwo}{266}, 1e-8 )
 ( \lumi{\thresholdopt}{\genevents}{\sigmatwo}{72}, 5e-9 )
 ( \lumi{\thresholdopt}{\geneventsbig}{\sigmatwo}{6}, 1e-9)
};
\addlegendentry{200}

\addplot+[smooth, plotthree, fillplot] coordinates{
 (  \lumi{\thresholdmin}{\genevents}{\sigmathree}{2561}, 5e-8 )
 (  \lumi{\thresholdmin}{\genevents}{\sigmathree}{555}, 1e-8 )
 (  \lumi{\thresholdmin}{\genevents}{\sigmathree}{139}, 5e-9 )
 (  \lumi{\thresholdmin}{\geneventsbig}{\sigmathree}{20}, 1e-9 )
 (  \lumi{\thresholdmax}{\geneventsbig}{\sigmathree}{20}, 1e-9 )
 (  \lumi{\thresholdmax}{\genevents}{\sigmathree}{139}, 5e-9 )
 (  \lumi{\thresholdmax}{\genevents}{\sigmathree}{555}, 1e-8 )
 (  \lumi{\thresholdmax}{\genevents}{\sigmathree}{2561}, 5e-8 )
} \closedcycle;

\addplot+[smooth, plotthree] coordinates{
 (  \lumi{\thresholdopt}{\genevents}{\sigmathree}{2561}, 5e-8 )
 (  \lumi{\thresholdopt}{\genevents}{\sigmathree}{555}, 1e-8 )
 (  \lumi{\thresholdopt}{\genevents}{\sigmathree}{139}, 5e-9 )
 (  \lumi{\thresholdopt}{\geneventsbig}{\sigmathree}{20}, 1e-9 )
};
\addlegendentry{300}

\addplot+[smooth, plotfour, fillplot] coordinates{
 (  \lumi{\thresholdmin}{\genevents}{\sigmafour}{2274}, 5e-8 )
 ( \lumi{\thresholdmin}{\genevents}{\sigmafour}{757}, 1e-8 )
 ( \lumi{\thresholdmin}{\genevents}{\sigmafour}{204}, 5e-9 )
 ( \lumi{\thresholdmin}{\geneventsbig}{\sigmafour}{20}, 1e-9 )
 ( \lumi{\thresholdmax}{\geneventsbig}{\sigmafour}{20}, 1e-9 )
 ( \lumi{\thresholdmax}{\genevents}{\sigmafour}{204}, 5e-9 )
 ( \lumi{\thresholdmax}{\genevents}{\sigmafour}{757}, 1e-8 )
 (  \lumi{\thresholdmax}{\genevents}{\sigmafour}{2274}, 5e-8 )
} \closedcycle;

\addplot+[smooth, plotfour] coordinates{
 (  \lumi{\thresholdopt}{\genevents}{\sigmafour}{2274}, 5e-8 )
 ( \lumi{\thresholdopt}{\genevents}{\sigmafour}{757}, 1e-8 )
 ( \lumi{\thresholdopt}{\genevents}{\sigmafour}{204}, 5e-9 )
 ( \lumi{\thresholdopt}{\geneventsbig}{\sigmafour}{20}, 1e-9 )
};
\addlegendentry{400}

\addplot [ pattern = north east lines ]
coordinates{
 ( \xmin, \fermibound )
 ( \xmax, \fermibound )
 ( \xmax, \ymax )
 ( \xmin, \ymax )
} \closedcycle ;

\node [fill = white] at ( axis cs: 100, 1.5e-8) {Fermi-LAT Excluded};

\end{loglogaxis}
\end{tikzpicture}
\caption{%
Mass reconstruction reach at \unit[8]{TeV} under the assumption that $S = 40$ events (middle line of each colored band) are sufficient in order to reconstruct the neutralino mass.
The lower and bands correspond to $S=30$ and $S=50$ events, respectively, and $P_\text{obs} \approx \unit[50]{\%}$.%
}
\label{fig:mass reconstruction reach}
\end{figure}
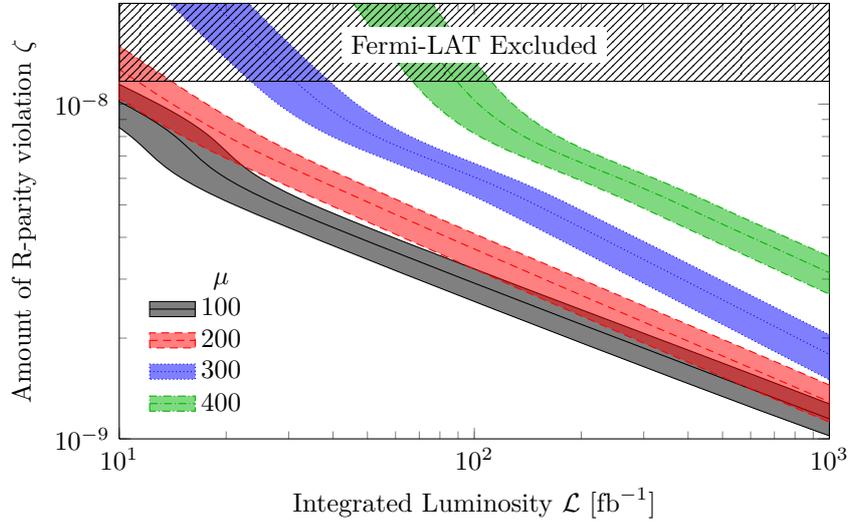

After applying the mass edge method described in section~\ref{sec:mass determination} to one of our benchmark models, we estimate that 30 signal events are sufficient to reconstruct the neutralino mass with a couple of GeV's precision.
Figure~\ref{fig:mass reconstruction reach} shows the resulting integrated luminosities at which the higgsino mass could be determined at the LHC running at \unit[8]{TeV} for our benchmark models.

\begin{figure}
\centering
\small
\tikzsetnextfilename{Discovery-Reach-14}
\newcommand{\xmin}{1}
\newcommand{\xmax}{100}
\newcommand{\ymin}{1e-9}
\newcommand{\ymax}{2e-8}

\newcommand{\sigmaone}{7649}
\newcommand{\sigmatwo}{532.6}
\newcommand{\sigmathree}{96.47}
\newcommand{\sigmafour}{25.73}

\newcommand{\genevents}{50000}
\newcommand{\geneventsbig}{140000}

\newcommand{\thresholdmin}{5.67}
\newcommand{\thresholdopt}{9.27}
\newcommand{\thresholdmax}{13.1}
\newcommand{\fermibound}{1.17e-8}

\newcommand{\lumi}[4]{ ( #1 *#2 ) / ( #3 * #4 ) }  

\pgfplotsset{
  legend image code/.code={%
    \draw[#1, fill, fill opacity = 0.5] (0cm,-0.1cm) rectangle (0.6cm,0.1cm);
    \draw[#1] (0cm,0cm) -- (0.6cm,00cm);
  }
}

\pgfplotsset{
 fillplot/.style = { fill, fill opacity = 0.5, forget plot }
}

\begin{tikzpicture}

\pgfplotstableread{lum14_fb_Pobs50_mu100.dat}\minone
\pgfplotstableread{lum14_fb_Pobs90_mu100.dat}\medone
\pgfplotstableread{lum14_fb_Pobs99_mu100.dat}\maxone

\pgfplotstableread{lum14_fb_Pobs50_mu200.dat}\mintwo
\pgfplotstableread{lum14_fb_Pobs90_mu200.dat}\medtwo
\pgfplotstableread{lum14_fb_Pobs99_mu200.dat}\maxtwo

\pgfplotstableread{lum14_fb_Pobs50_mu300.dat}\minthree
\pgfplotstableread{lum14_fb_Pobs90_mu300.dat}\medthree
\pgfplotstableread{lum14_fb_Pobs99_mu300.dat}\maxthree

\pgfplotstableread{lum14_fb_Pobs50_mu400.dat}\minfour
\pgfplotstableread{lum14_fb_Pobs90_mu400.dat}\medfour
\pgfplotstableread{lum14_fb_Pobs99_mu400.dat}\maxfour

\pgfplotstablesort[sort key={zeta},sort cmp={float >}]\minonesort{\minone}
\pgfplotstablesort[sort key={zeta},sort cmp={float <}]\maxonesort{\maxone}

\pgfplotstablesort[sort key={zeta},sort cmp={float >}]\mintwosort{\mintwo}
\pgfplotstablesort[sort key={zeta},sort cmp={float <}]\maxtwosort{\maxtwo}

\pgfplotstablesort[sort key={zeta},sort cmp={float >}]\minthreesort{\minthree}
\pgfplotstablesort[sort key={zeta},sort cmp={float <}]\maxthreesort{\maxthree}

\pgfplotstablesort[sort key={zeta},sort cmp={float >}]\minfoursort{\minfour}
\pgfplotstablesort[sort key={zeta},sort cmp={float <}]\maxfoursort{\maxfour}

\pgfplotstablevertcat{\fillone}{\minonesort}
\pgfplotstablevertcat{\fillone}{\maxonesort}

\pgfplotstablevertcat{\filltwo}{\mintwosort}
\pgfplotstablevertcat{\filltwo}{\maxtwosort}

\pgfplotstablevertcat{\fillthree}{\minthreesort}
\pgfplotstablevertcat{\fillthree}{\maxthreesort}

\pgfplotstablevertcat{\fillfour}{\minfoursort}
\pgfplotstablevertcat{\fillfour}{\maxfoursort}

\begin{loglogaxis}[
      xmin = \xmin
 , xmax = \xmax
 , xlabel = {Integrated Luminosity $\mathcal L $ [$\text{fb}^{-1}$] }
 , ymin = \ymin
 , ymax = \ymax
 , ylabel = {Amount of R-parity violation $\zeta$}
 , extra y tick style={ log identify minor tick positions = true }
 , legend style = { legend pos = south west }
]

\addlegendimage{empty legend}
\addlegendentry{$\mu$}

\addplot[ plotone, fill, fill opacity=.5, smooth, forget plot]  table [ x index = 0, y index = 1 ] {\fillone};
\addplot[plotone, smooth] table[ x expr = \thisrow{Lum}, y expr = \thisrow{zeta}]  {\medone};
\addlegendentry{100}

\addplot[ plottwo, fill, fill opacity=.5, smooth , forget plot]  table [ x index = 1, y index = 0 ] {\filltwo};
\addplot[plottwo, smooth] table[ x expr = \thisrow{Lum}, y expr = \thisrow{zeta}]  {\medtwo};
\addlegendentry{200}

\addplot[ plotthree, fill, fill opacity=.5, smooth, forget plot ]  table [ x index = 0, y index = 1 ] {\fillthree};
\addplot[plotthree, smooth] table[ x expr = \thisrow{Lum}, y expr = \thisrow{zeta}]  {\medthree};
\addlegendentry{300}

\addplot[ plotfour, fill, fill opacity=.5, smooth, forget plot ]  table [ x index = 0, y index = 1 ] {\fillfour};
\addplot[plotfour, smooth] table[ x expr = \thisrow{Lum}, y expr = \thisrow{zeta}]  {\medfour};
\addlegendentry{400}

\addplot [ pattern = north east lines ]
coordinates{
 ( \xmin, \fermibound )
 ( \xmax, \fermibound )
 ( \xmax, \ymax )
 ( \xmin, \ymax )
} \closedcycle ;

\node [fill = white] at ( axis cs: 10, 1.5e-8) {Fermi-LAT Excluded};

\end{loglogaxis}

\end{tikzpicture}
\caption{%
Estimation of the \unit[14]{TeV} discovery reach based on our \unit[8]{TeV} results.
Each colored band represents a value of $\mu$ and the lower, middle and upper line on each band corresponds to $P_\text{obs} = \unit[50]{\%}$, \unit[90]{\%} and \unit[99]{\%}, respectively.}
\label{fig:Discovery Reach 14}
\end{figure}
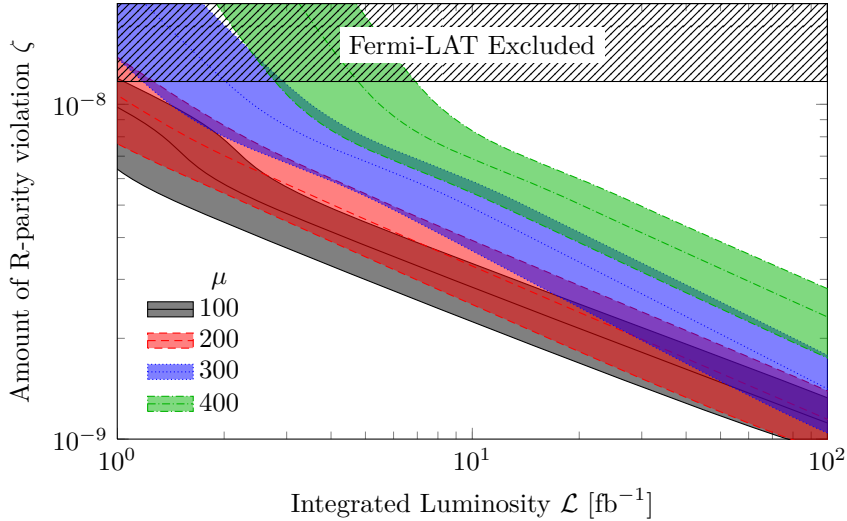%
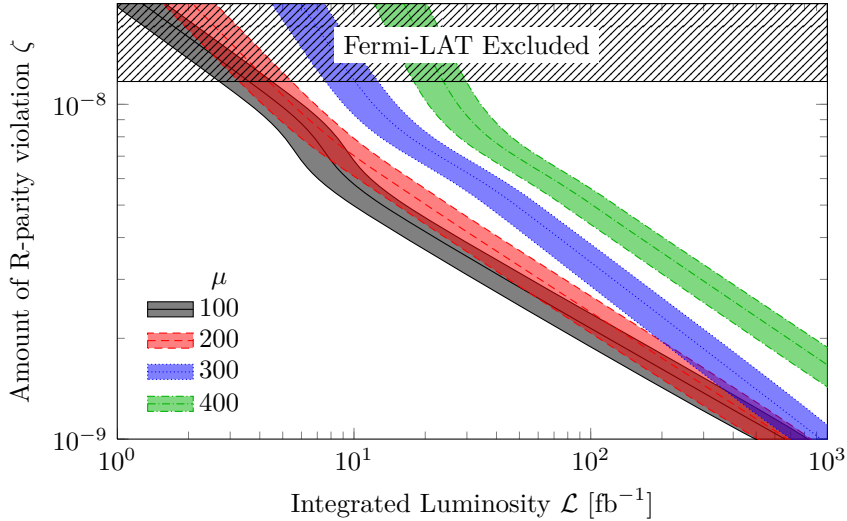
\begin{figure}
\centering
\small
\tikzsetnextfilename{Mass-Reconstruction-14}
\newcommand{\xmin}{1}
\newcommand{\xmax}{1000}
\newcommand{\ymin}{1e-9}
\newcommand{\ymax}{2e-8}
\newcommand{\sigmaone}{16320}
\newcommand{\sigmatwo}{1371}
\newcommand{\sigmathree}{298.4}
\newcommand{\sigmafour}{96.26}
\newcommand{\genevents}{50000}
\newcommand{\geneventsbig}{140000}
\newcommand{\geneventslarge}{4320000}
\newcommand{\thresholdmin}{30}
\newcommand{\thresholdopt}{40}
\newcommand{\thresholdmax}{50}
\newcommand{\fermibound}{1.17e-8}
\newcommand{\lumi}[4]{ ( #1 *#2 ) / ( #3 * #4 ) }  
\pgfplotsset{
 fillplot/.style = { fill, fill opacity = 0.5, forget plot }
}
\pgfplotsset{
  legend image code/.code={%
    \draw[#1, fill, fill opacity = 0.5] (0cm,-0.1cm) rectangle (0.6cm,0.1cm);
    \draw[#1] (0cm,0cm) -- (0.6cm,00cm);
  }
}
\begin{tikzpicture}
\begin{loglogaxis}[ 
  xmin = \xmin
, xmax = \xmax
, xlabel = { Integrated Luminosity $\mathcal L $ [$\text{fb}^{-1}$]  }
, ymin = \ymin
, ymax = \ymax
, ylabel = { Amount of R-parity violation $\zeta$ }
, extra y tick style={ log identify minor tick positions = true }
, legend style = { legend pos = south west }
]
\addlegendimage{empty legend}
\addlegendentry{$\mu$}

\addplot+[ smooth, plotone, fillplot]
coordinates{
 ( \lumi{\thresholdmin}{\geneventsbig}{\sigmaone}{1647}, 5e-8 )
 ( \lumi{\thresholdmin}{\geneventsbig}{\sigmaone}{71}, 1e-8 ) 
 ( \lumi{\thresholdmin}{\geneventsbig}{\sigmaone}{26}, 5e-9 ) 
 ( \lumi{\thresholdmin}{\geneventslarge}{\sigmaone}{16}, 1e-9 )
 ( \lumi{\thresholdmax}{\geneventslarge}{\sigmaone}{16}, 1e-9 )
 ( \lumi{\thresholdmax}{\geneventsbig}{\sigmaone}{26}, 5e-9 ) 
 ( \lumi{\thresholdmax}{\geneventsbig}{\sigmaone}{71}, 1e-8 )
 ( \lumi{\thresholdmax}{\geneventsbig}{\sigmaone}{1647}, 5e-8 )
} \closedcycle;

\addplot+[ smooth, plotone]
coordinates{
 ( \lumi{\thresholdopt}{\geneventsbig}{\sigmaone}{1647}, 5e-8 )
 ( \lumi{\thresholdopt}{\geneventsbig}{\sigmaone}{71}, 1e-8 )
 ( \lumi{\thresholdopt}{\geneventsbig}{\sigmaone}{26}, 5e-9 )
 ( \lumi{\thresholdopt}{\geneventslarge}{\sigmaone}{16}, 1e-9 )
};
\addlegendentry{100}

\addplot+[smooth, plottwo, fillplot ] coordinates{
 ( \lumi{\thresholdmin}{\genevents}{\sigmatwo}{2406}, 5e-8 )
 ( \lumi{\thresholdmin}{\genevents}{\sigmatwo}{266}, 1e-8 )
 ( \lumi{\thresholdmin}{\genevents}{\sigmatwo}{72}, 5e-9 )
 ( \lumi{\thresholdmin}{\geneventsbig}{\sigmatwo}{6}, 1e-9)
 ( \lumi{\thresholdmax}{\geneventsbig}{\sigmatwo}{6}, 1e-9)
 ( \lumi{\thresholdmax}{\genevents}{\sigmatwo}{72}, 5e-9 )
 ( \lumi{\thresholdmax}{\genevents}{\sigmatwo}{266}, 1e-8 )
 ( \lumi{\thresholdmax}{\genevents}{\sigmatwo}{2406}, 5e-8 )
} \closedcycle;

\addplot+[smooth, plottwo ] coordinates{
 ( \lumi{\thresholdopt}{\genevents}{\sigmatwo}{2406}, 5e-8 )
 ( \lumi{\thresholdopt}{\genevents}{\sigmatwo}{266}, 1e-8 )
 ( \lumi{\thresholdopt}{\genevents}{\sigmatwo}{72}, 5e-9 )
 ( \lumi{\thresholdopt}{\geneventsbig}{\sigmatwo}{6}, 1e-9)
};
\addlegendentry{200}

\addplot+[smooth, plotthree, fillplot] coordinates{
 (  \lumi{\thresholdmin}{\genevents}{\sigmathree}{2561}, 5e-8 )
 (  \lumi{\thresholdmin}{\genevents}{\sigmathree}{555}, 1e-8 )
 (  \lumi{\thresholdmin}{\genevents}{\sigmathree}{139}, 5e-9 )
 (  \lumi{\thresholdmin}{\geneventsbig}{\sigmathree}{20}, 1e-9 )
 (  \lumi{\thresholdmax}{\geneventsbig}{\sigmathree}{20}, 1e-9 )
 (  \lumi{\thresholdmax}{\genevents}{\sigmathree}{139}, 5e-9 )
 (  \lumi{\thresholdmax}{\genevents}{\sigmathree}{555}, 1e-8 )
 (  \lumi{\thresholdmax}{\genevents}{\sigmathree}{2561}, 5e-8 )
} \closedcycle;

\addplot+[smooth, plotthree] coordinates{
 (  \lumi{\thresholdopt}{\genevents}{\sigmathree}{2561}, 5e-8 )
 (  \lumi{\thresholdopt}{\genevents}{\sigmathree}{555}, 1e-8 )
 (  \lumi{\thresholdopt}{\genevents}{\sigmathree}{139}, 5e-9 )
 (  \lumi{\thresholdopt}{\geneventsbig}{\sigmathree}{20}, 1e-9 )
};
\addlegendentry{300}

\addplot+[smooth, plotfour, fillplot] coordinates{
 (  \lumi{\thresholdmin}{\genevents}{\sigmafour}{2274}, 5e-8 )
 ( \lumi{\thresholdmin}{\genevents}{\sigmafour}{757}, 1e-8 )
 ( \lumi{\thresholdmin}{\genevents}{\sigmafour}{204}, 5e-9 )
 ( \lumi{\thresholdmin}{\geneventsbig}{\sigmafour}{20}, 1e-9 )
 ( \lumi{\thresholdmax}{\geneventsbig}{\sigmafour}{20}, 1e-9 )
 ( \lumi{\thresholdmax}{\genevents}{\sigmafour}{204}, 5e-9 )
 ( \lumi{\thresholdmax}{\genevents}{\sigmafour}{757}, 1e-8 )
 (  \lumi{\thresholdmax}{\genevents}{\sigmafour}{2274}, 5e-8 )
} \closedcycle;

\addplot+[smooth, plotfour] coordinates{
 (  \lumi{\thresholdopt}{\genevents}{\sigmafour}{2274}, 5e-8 )
 ( \lumi{\thresholdopt}{\genevents}{\sigmafour}{757}, 1e-8 )
 ( \lumi{\thresholdopt}{\genevents}{\sigmafour}{204}, 5e-9 )
 ( \lumi{\thresholdopt}{\geneventsbig}{\sigmafour}{20}, 1e-9 )
};
\addlegendentry{400}

\addplot [ pattern = north east lines ]
coordinates{
 ( \xmin, \fermibound )
 ( \xmax, \fermibound )
 ( \xmax, \ymax )
 ( \xmin, \ymax )
} \closedcycle ;

\node [fill = white] at ( axis cs: 30, 1.5e-8) {Fermi-LAT Excluded};

\end{loglogaxis}
\end{tikzpicture}
\caption{%
Estimation of the mass reconstruction reach for \unit[14]{TeV} based on our \unit[8]{TeV} analysis, under the assumption that $S = 40$ events are sufficient in order to reconstruct the neutralino mass.
The bands correspond to the interval spanning between 30 and 50 events, and $P_\text{obs} \approx \unit[50]{\%}$.%
}
\label{fig:mass reconstruction reach 14}
\end{figure}

We use our \unit[8]{TeV} results to estimate the reach when LHC runs at the design centre-of-mass energy of \unit[14]{TeV}, by the same statistical analysis applied after a naive scaling of the cross sections after cuts with the factors $\nicefrac{ \sigma_{14}^\text{tot} }{ \sigma_8^\text{tot} }$ presented in Table~\ref{tab:prodxsec}.
Since the background is assumed to be completely removed by our cuts, the reach at \unit[14]{TeV} would be significantly improved.
Larger higgsino masses can be reached at smaller integrated luminosities, as can be observed in Figure~\ref{fig:Discovery Reach 14}.
Also the luminosity which is needed in order to reconstruct the neutralino masses is reduced, as can be seen in Figure~\ref{fig:mass reconstruction reach 14}.
These results are approximate and we expect that changes in \emph{e.g.}\ the $p_T$ cut on the muons may be needed to deal with systematic effects such as increasing pileup.

\section{Conclusion}
\label{sec:conclusion}

We have investigated the LHC detection prospects for the light-higgsino scenario, or LHS, in the MSSM extended with bilinear R-parity breaking terms.
A spectrum with the higgsino-like neutralinos and chargino light, of the order of \unit[100]{GeV}, and the other superparticles in the multi-TeV range is consistent with a Higgs mass of $\sim \unit[126]{GeV}$ within the MSSM and can be obtained in GUT models.
Because the higgsinos are nearly mass degenerate and the strongly interacting superparticles are out of reach, such a scenario within the usual MSSM is difficult to probe at the LHC.
The prospects change if we allow for R-parity violation, which leads to a consistent cosmology where leptogenesis and gravitino dark matter can be accounted for without conflict with BBN.

We calculated the R-parity violating decay modes for a higgsino-like NLSP, and thereby diagonalized the full R-parity breaking neutralino matrix for this case.
We showed that neither the resulting mass eigenstates nor the transformation matrices differ from the ones derived previously in the bino NLSP case.
In fact, the parametrization only depends on a large hierarchy between the fermi scale and the largest neutralino mass.

We were thus able to use the limits on decaying gravitino dark matter from gamma-ray searches with the Fermi-LAT to put an upper bound on the RPV parameter $\zeta$ and thereby a lower bound on the higgsino NLSP decay length.
This motivated our study of a dimuon LHC signature with macroscopic and, compared to other scenarios studied in the literature, relatively large displaced vertices.
We simulated events and detector response for a few benchmark models, varying the value of the higgsino mass parameter $\mu$ and the RPV parameter $\zeta$.
We found that this scenario can show up already in the \unit[8]{TeV} run at the LHC with the $\unit[10 \text{ -- }30]{fb^{-1}}$ of data expected at this center-of-mass energy, and that the reach in $\zeta$ is improved by up to an order of magnitude compared to the current reach of gamma-ray searches.

We also demonstrated that in the case of a signal, the LHS with RPV hypothesis can be tested further by a mass edge reconstruction.
Except for the largest $\zeta$ and smallest higgsino masses considered, the luminosity accumulated during the \unit[8]{TeV} run may not be enough.
However, based on a our rough estimation of the \unit[14]{TeV} reach, the higgsino mass in all our benchmark scenarios can be determined in the \unit[14]{TeV} run, requiring integrated luminosities in the range $\unit[30 \text{ -- } 1000]{fb^{-1}}$.

The complementarity of gamma-ray and LHC signatures for the LHS with RPV may also be used to falsify the model. A future observation of a gamma-ray line consistent with decaying dark matter would fix the gravitino mass and the value of $\zeta$. If also an LHC signal of a higgsino-like neutralino shows up, which may even be possible in the scenario considered here if the stops are light enough to be produced at the higher center-of-mass energy,  the absence of any displaced decays would then render the LHS with RPV in conflict with experiments.

\appendix

\section{Higgsino and gravitino branching ratios}
\label{App:BR}

The couplings of the gauge fields to charged and neutral matter are given by
\begin{align}
    \mathcal L
  = - e J_{ e \mu } A^\mu
  - \frac{ g }{ c_w } J_{ Z \mu } Z^\mu
  - \frac{ g }{ \sqrt 2 } J_\mu^- W^{ + \mu }
  - \frac{ g }{ \sqrt 2 } J_\mu^+ W^{ - \mu }
\ .
\end{align}
The currents in the gauge eigenstate basis are
\begin{subequations}
\begin{align}
    J_{ e \mu }
 =& J^3_{e \mu} + J^{2, 1}_{e \mu} \\
=&\ \overline w^+ \overline \sigma_\mu w^+
  - \overline w^- \overline \sigma_\mu w^-
  - \overline e_i \overline \sigma_\mu e_i
  + \overline e_i^c \overline \sigma_\mu e_i^c
  - \overline h^-_d \overline \sigma_\mu h^-_d
  + \overline h^+_u \overline \sigma_\mu h^+_u
\ , \notag \\
    J_{ Z \mu }
=&- \frac{ 1 }{ 2 } \overline h_u^0 \overline \sigma_\mu h_u^0
  + \frac{ 1 }{ 2 } \overline h_d^0 \overline \sigma_\mu h_d^0
  + \frac{ 1 }{ 2 } \overline \nu_i \overline \sigma_\mu \nu_i \\
 &+ \overline w^+ \overline \sigma_\mu w^+
  - \overline w^- \overline \sigma_\mu w^-
  - \frac{ 1 }{ 2 } \overline e_i \overline \sigma_\mu e_i
  - \frac{ 1 }{ 2 } \overline h_d^- \overline \sigma_\mu h_d^-
  + \frac{ 1 }{ 2 } \overline h_u^+ \overline \sigma_\mu h_u^+
  - s_w^2 J^{2,1}_{ e \mu }
\ , \notag \\
    J_\mu^-
=&\ \sqrt 2
    \left( \overline w^3 \overline \sigma_\mu w^-
      - \overline w^+ \overline \sigma_\mu w^3 \right)
  + \overline \nu_i \overline \sigma_\mu e_i
  + \overline h^0_d \overline \sigma_\mu h^-_d
  + \overline h^+_u \overline \sigma_\mu h^0_u
\ , \\
    J_\mu^+
=&\ \sqrt{2}
    \left( \overline w^- \overline \sigma_\mu w^3
      - \overline w^3 \overline \sigma_\mu w^+ \right)
  - \overline e_i \overline \sigma_\mu \nu_i
  - \overline h^-_d \overline \sigma_\mu h^0_d
  - \overline h^0_u \overline \sigma_\mu h^+_u
\ .
\end{align}
\end{subequations}
The upper indices of the electromagnetic currents indicate  the transformation properties of the fields in the current under the $SU(2)_\text{L}$.
Having derived the currents we now have to transform them into the mass-eigenstate basis of the fermions.
To this end, we have to diagonalize the mass matrices $\mathcal M^N$ and $\mathcal M^C$, which in the gauge eigenbasis are%
\footnote{Note the extra factors of $\sqrt 2$ in the charged mass matrix compared to~\cite{Bobrovskyi:2010ps}.}
\begin{subequations}
\begin{align}
    \mathcal M^N
 &= \begin{pmatrix}
       M_1
     & 0
     & m_Z s_\beta s_w
     & - m_Z c_\beta s_w
     & - \zeta_i m_Z s_w
    \\ 0
     & M_2
     & - m_Z s_\beta c_w
     & m_Z c_\beta c_w
     &  \zeta_i m_Z c_w
    \\ m_Z s_\beta s_w
     & - m_Z s_\beta c_w
     & 0
     & -\mu
     & 0
    \\ - m_Z c_\beta s_w
     & m_Z c_\beta c_w
     & - \mu
     & 0
     & 0
    \\ - \zeta_i m_Z s_w
     & \zeta_i m_Z c_w
     & 0
     & 0
     & 0
    \end{pmatrix}
    \label{neutralino mixing matrix}
\ , \\
    \mathcal M^C
 &= \begin{pmatrix}
       M_2
     & \sqrt 2 m_Z s_\beta c_w
     & 0
    \\ \sqrt 2 m_Z c_\beta c_w
     & \mu
     & \zeta_i h^e_{ i j } \mu
    \\ \sqrt 2 \zeta_i m_Z c_w
     & 0
     & h^e_{ i j } v c_\beta
   \end{pmatrix}
\ , \qquad h_{ij}^e = \diag( h_1^e, h_2^e, h_3^e )
\ .
    \label{chargino mixing matrix}
\end{align}
\end{subequations}
The matrices $\mathcal M^N$ and $\mathcal M^C$ are diagonalized by unitary and bi-unitary transformations, respectively,
\begin{align}
    U^{ ( n ) T } \mathcal M^N U^{ ( n ) }
   = \mathcal M^N_\text{diag}
\ , &&
    U^{ ( c ) \dagger } \mathcal M^C \widetilde U^{ ( c ) }
   = \mathcal M^C_\text{diag}
\ ,
\end{align}
where
$U^{ ( n ) \dagger } U^{ ( n ) } = U^{ ( c ) \dagger } U^{ ( c ) } = \widetilde{ U }^{ ( c ) \dagger } \widetilde{ U }^{ ( c ) } = \mathbf 1$.
These unitary transformations relate the neutral and charged gauge eigenstates to the mass eigenstates $( \chi_a^0, \nu^\prime_i )$ ($ a = 1, \ldots, 4 $) and $( \chi^-_\alpha, e'_i )$, $( \chi^+_\alpha, e^{ \prime c }_i )$ ($\alpha = 1, 2$), respectively,
\begin{align}
    \begin{pmatrix}
       b
    \\ w^3
    \\ h^0_u
    \\ h^0_d
    \\ \nu_i
    \end{pmatrix}
  = U^{ ( n ) }
    \begin{pmatrix}
       \chi^0_1
    \\ \chi^0_2
    \\ \chi^0_3
    \\ \chi^0_4
    \\ \nu_i^\prime
    \end{pmatrix}
\ , &&
    \begin{pmatrix}
       w^-
    \\ h_d^-
    \\ e_i
    \end{pmatrix}
  = U^{ ( c ) }
    \begin{pmatrix}
       \chi^-_1
    \\ \chi^-_2
    \\ e^\prime_i
    \end{pmatrix}
\ , &&
    \begin{pmatrix}
       w^+
    \\ h_u^+
    \\ e^c_i
    \end{pmatrix}
  = \widetilde U^{ ( c ) }
    \begin{pmatrix}
       \chi^+_1
    \\ \chi^+_2
    \\ e^{ c \prime }_i
    \end{pmatrix}
\ .
\end{align}
Our convention for the separation of the transformation matrices into R-parity conserving and violating parts is
\begin{subequations}
\setlength{\extrarowheight}{1ex}
\begin{align}
    U^{ ( n ) }
  &= \left ( \begin{array}{c|c}
      U_{ a b }^{ ( \chi^0 ) }
    & U_{ a i }^{ ( \chi^0, \nu ) } \\ \hline
      U_{ i a }^{ ( \nu, \chi^0 ) }
    & U_{ i j }^{ ( \nu ) }
    \end{array} \right)
\ , \\
    U^{ ( c ) }
  &= \left ( \begin{array}{c|c}
      U_{ \alpha \beta }^{ ( \chi^- ) }
    & U_{ \alpha i }^{ ( \chi^-, e ) } \\[0.3ex] \hline
      U_{ i \alpha }^{ ( e, \chi^- ) }
    & U_{ i j }^{ ( e ) }
    \end{array} \right)
\ , &&
    \widetilde U^{ ( c ) }
  = \left ( \begin{array}{c|c}
      \widetilde U_{ \alpha \beta }^{ ( \chi^+ ) }
    & \widetilde U_{ \alpha i }^{ ( \chi^+, e^c ) } \\[0.3ex] \hline
      \widetilde U_{ i \alpha }^{ ( e^c, \chi^+ ) }
    & \widetilde U_{ i j }^{ ( e^c ) }
    \end{array} \right)
\ .
\end{align}
\end{subequations}
After the transformation the currents can be expressed through CKM-like products of two transformation matrices
\begin{subequations}
\begin{align}
    J_{ Z \mu }
=&\ \overline \chi^0_a \overline \sigma_\mu V^{ ( \chi^0 ) }_{ a b } \chi^0_b
  + \overline \chi^-_\alpha \overline \sigma_\mu V^{ ( \chi^- ) }_{ \alpha \beta } \chi^-_\beta
  + \overline \chi^+_\alpha \overline \sigma_\mu V^{ ( \chi^+ ) }_{ \alpha \beta } \chi^+_\beta \\
 &+ \overline \nu_i \overline \sigma_\mu V^{ ( \nu ) }_{ i j } \nu_j
  + \overline e_i \overline \sigma_\mu V^{ ( e  ) }_{ i j } e_j
  + \overline e_i^c \overline \sigma_\mu V^{ ( e^c  ) }_{ i j } e_j^c \notag \\
 &+ \left( \overline \chi^0_a \overline \sigma_\mu V^{ ( \chi, \nu ) }_{ a j } \nu_j
    + \overline \chi^-_\alpha \overline \sigma_\mu V^{ ( \chi^-, e ) }_{ \alpha j } e_j
    + \overline \chi^+_\alpha \overline \sigma_\mu V^{ ( \chi^+, e^c ) }_{ \alpha j } e_j^c
    + \text{h.c.} \right)
  - s_w^2 J_{ e \mu } \notag
\ ,
 \\ J_{ e \mu }
=&\ \overline \chi^-_\alpha \overline \sigma_\mu V^{ ( \chi^- ) }_{ \alpha \beta } \chi^-_\beta
  + \overline \chi^+_\alpha \overline \sigma_\mu V^{ ( \chi^+ ) }_{ \alpha \beta } \chi^+_\beta
  + \overline e_i \overline \sigma_\mu V^{ ( e  ) }_{ i j } e_j
  + \overline e_i^c \overline \sigma_\mu V^{ ( e^c  ) }_{ i j } e_j^c \\
& + \left( \overline \chi^-_\alpha \overline \sigma_\mu V^{ ( \chi^-, e ) }_{ \alpha j } e_j
    + \overline \chi^+_\alpha \overline \sigma_\mu V^{ ( \chi^+, e^c ) }_{ \alpha j } e_j^c
    + \text{h.c.} \right)
\ , \notag
\end{align}
\end{subequations}
where the CKM-like matrices relevant for this study are given by
\begin{subequations}
\begin{align}
    V_{ a j }^{ ( \chi, \nu ) }
 &= - \frac{ 1 }{ 2 } U_{ 3 a }^{ ( \chi^0 ) } U_{ 3 j }^{ ( \chi^0, \nu ) }
  + \frac{ 1 }{ 2 } U_{ 4 a }^{ ( \chi^0 ) } U_{ 4 j }^{ ( \chi^0, \nu ) }
  + \frac{ 1 }{ 2 } \sum_k U_{ k a }^{ ( \nu, \chi^0 ) } U_{ k j }^{ ( \nu ) }
\ , 
 \\ V_{ \alpha j }^{ ( \chi^-, e ) }
 &= \left( s_w^2 - 1 \right) U_{ 1 \alpha }^{ ( \chi^- ) } U_{ 1 j }^{ ( \chi^-, e ) }
    + \left( s_w^2 - \frac{ 1 }{ 2 } \right) \left( U_{ 2 \alpha }^{ ( \chi^- ) } U_{ 2 j }^{ ( \chi^-, e ) }
    + 
        \sum_i U_{ i \alpha }^{ ( e, \chi^- ) } U_{ i j }^{ ( e ) } \right)
\ .
\end{align}
\end{subequations}
The following results are approximated by an expansion in $\zeta$ and $\epsilon$, where $\epsilon = \nicefrac{ m_Z }{ \tilde m }$ and $\widetilde m$ the largest mass parameter of either $M_1$, $M_2$ or $\mu$.
The parameter choice affects neither the expansion of $U^{ ( n ) }$ nor the mass eigenstates.
For the R-parity violating part of the neutral CKM-like matrix we find
\begin{align}
    V^{ ( \chi, \nu ) }_{ a j }
=&\ - \frac{ 1 }{ 2 } \zeta_j m_Z
     \begin{pmatrix}
       \frac{ s_w }{ M_1 } \\
     - \frac{ c_w }{ M_2 } \\
       \frac{ m_Z }{ \sqrt 2 \mu } v_1 \\
       \frac{ m_Z }{ \sqrt 2 \mu } v_2
    \end{pmatrix}
    \left( 1 + \mathcal O
    \begin{pmatrix}[r]
      s_{ 2 \beta } \left( \frac{ m_Z }{ \tilde m } \right)^2 \\
      \left( \frac{ m_Z }{ \tilde m } \right)^2 \\
      \left( \frac{ m_Z }{ \tilde m } \right)^2 \\
      s_{ 2 \beta } \left( \frac{ m_Z }{ \tilde m } \right)^2
    \end{pmatrix} \right)
\ ,
\end{align}
with abbreviations
\begin{subequations}
\begin{align}
    v_1
 =& ( s_\beta + c_\beta ) \frac{ M_1 c_w^2 + M_2 s_w^2 - \mu }{ ( M_1 - \mu )( M_2 - \mu ) }
  - ( s_\beta - c_\beta ) \left( \frac{ s_w^2 }{ M_1 } + \frac{ c_w^2 }{ M_2 } \right)
\ , \\
    v_2
 =& ( s_\beta - c_\beta ) \frac{ M_1 c_w^2 + M_2 s_w^2 + \mu }{ ( M_1 + \mu )( M_2 + \mu ) }
  - ( s_\beta + c_\beta ) \left( \frac{ s_w^2 }{ M_1 } + \frac{ c_w^2 }{ M_2 } \right)
\ .
\end{align}
\end{subequations}
Numerically, the relative errors are smaller than $0.10$, $0.20$, $0.15$, $0.05$ for $a = 1, \dots, 4$.
For the R-parity violating part of the charged CKM-like matrix we find
\begin{align}
    V^{ ( \chi, e ) }_{ a j }
=&\ - \zeta_j m_Z
     \begin{pmatrix}
       \frac{ s_w }{ M_1 } \\
       \frac{ c_w }{ M_2 } \\
       \frac{ m_Z }{ \sqrt 2 \mu } \widetilde v_1 \\
       \frac{ m_Z }{ \sqrt 2 \mu } \widetilde v_2
    \end{pmatrix}
    \left( 1 + \mathcal O
    \begin{pmatrix}[r]
      s_{ 2 \beta } \left( \frac{ m_Z }{ \tilde m } \right)^2 \\
      \left( \frac{ m_Z }{ \tilde m } \right)^2 \\
      \left( \frac{ m_Z }{ \tilde m } \right)^2 \\
      s_{ 2 \beta } \left( \frac{ m_Z }{ \tilde m } \right)^2
    \end{pmatrix} \right)
\ ,
\end{align}
with abbreviations%
\footnote{The formula for $\widetilde v_1$ differs slightly from the version in~\cite{Bobrovskyi:2010ps}, due to a typo therein.}
\begin{subequations}
\begin{align}
    \widetilde v_1
 =& ( s_\beta + c_\beta ) \frac{ M_1 c_w^2 + M_2 s_w^2 - \mu }{ ( M_1 - \mu )( M_2 - \mu ) }
  - 2 ( s_\beta + c_\beta ) \frac{ \mu c_w^2 }{ M_2 ( M_2 - \mu ) }
  + 2 s_\beta \frac{ c_w^2 }{ M_2 }
\ ,  \\
    \widetilde v_2
 =& ( s_\beta - c_\beta ) \frac{ M_1 c_w^2 + M_2 s_w^2 + \mu }{ ( M_1 + \mu )( M_2 + \mu ) }
  - 2 ( s_\beta + \frac{ \mu }{ M_2 } c_\beta ) \frac{ M_1 + \mu }{ ( M_1 + \mu ) ( M_2 + \mu ) } c_w^2
\ .
\end{align}
\end{subequations}
Here we again neglected corrections that involve the Yukawa couplings $h_{ i i }^e$.
The numerical corrections to the NLO contributions to $V^{ ( \chi, e ) }_{ a i }$ are smaller than 0.05, 0.15, 0.20 for $a = 1$, 2, 3, respectively.
For $a = 4$ we reach the limit of our numerical precision.

\acknowledgments

The authors thank F.~Br\"ummer, W.~Buchm\"uller and S.~Strandberg for discussions.

\bibliographystyle{JHEP}
\bibliography{higgsino-nlsp-w-rpv}

\end{document}